%% file: ms.tex
\documentclass[fleqn,usenatbib]{mnras}
\usepackage{url}
\usepackage{natbib}
\usepackage[normalem]{ulem}
\usepackage[T1]{fontenc}
\usepackage{txfonts}
\usepackage{graphicx}
\usepackage{savesym}
\savesymbol{leftroot}
\savesymbol{uproot}
\savesymbol{iint}
\savesymbol{iiint}
\savesymbol{iiiint}
\savesymbol{idotsint}
\savesymbol{Bbbk}
\usepackage{amsmath,amssymb,physics,bm}
\usepackage{xlmacros}
\usepackage{cancel}
\usepackage{amsmath,esint}
\usepackage[pdfpagelabels=false]{hyperref}
\hypersetup{colorlinks=True,linkcolor={RoyalBlue},citecolor={RoyalBlue},urlcolor={BrickRed},breaklinks=true}

\DeclareRobustCommand{\VAN}[3]{#2}
\let\VANthebibliography\thebibliography
\def\thebibliography{\DeclareRobustCommand{\VAN}[3]{##3}\VANthebibliography}

\title[FPFS v2]{Weak gravitational lensing shear measurement with FPFS:\\
    analytical mitigation of noise bias and selection bias}
\author[]{
    Xiangchong Li $^{1,2,3}$\thanks{xiangchl@andrew.cmu.edu},
    Yin Li  $^{4,1}$,
    Richard Massey $^{5}$\\
$^{1}$Kavli Institute for the Physics and Mathematics of the Universe (Kavli
IPMU, WPI), UTIAS, The University of Tokyo, 5-1-5 Kashiwanoha, Kashiwa,\\
Chiba 277-8583, Japan\\
$^{2}$Department of Physics, The University of Tokyo, 7-3-1 Hongo, Bunkyo,
Tokyo 113-0033, Japan\\
$^{3}$Department of Physics, McWilliams Center for Cosmology, Carnegie Mellon
University, Pittsburgh, PA 15213, USA\\
$^{4}$Center for Computational Astrophysics \& Center for Computational
Mathematics, Flatiron Institute, 162 5th Avenue, 10010, New York, NY, USA\\
$^{5}$Department of Physics, Institute for Computational Cosmology,
Durham University, South Road, Durham, DH1 3LE, UK \vspace{-4mm}
}
\date{Received Month XX, YYYY; accepted Month XX, YYYY}
\pubyear{2021}

\begin{document}
\label{firstpage}
\pagerange{\pageref{firstpage}--\pageref{lastpage}}
\maketitle

\begin{abstract}
    Dedicated `Stage IV' observatories will soon observe the entire
    extragalactic sky, to measure the `cosmic shear' distortion of galaxy
    shapes by weak gravitational lensing.
    To measure the apparent shapes of those galaxies, we present an improved
    version of the Fourier Power Function Shapelets (\FPFS{}) shear measurement
    method.  This now includes analytic corrections for sources of bias that
    plague all shape measurement algorithms: including noise bias (due to noise
    in nonlinear combinations of observable quantities) and selection bias (due
    to sheared galaxies being more or less likely to be detected).  Crucially,
    these analytic solutions do not rely on calibration from external image
    simulations.
    For isolated galaxies, the small residual $\sim$$10^{-3}$ multiplicative
    bias and $\lesssim$$10^{-4}$ additive bias now meet science requirements
    for Stage IV experiments.
    \xlrv{\FPFS{} also works accurately for faint galaxies and robustly against
    stellar contamination.} Future work will focus on deblending overlapping
    galaxies. The code used for this paper can process $>$$1000$ galaxy images
    per CPU~second and is available from
    \url{https://github.com/mr-superonion/FPFS}.
\end{abstract}


\section{INTRODUCTION}
\label{sec:Intro}

The images of distant galaxies appear weakly but coherently distorted because
light from them is deflected by the gravity of intervening matter along our
line of sight \citep[for reviews, see][]{WL-rev-Bartelmann01,
cosmicShear-rev-Kilbinger15}. The anisotropic stretch of galaxy images is
termed weak gravitational lensing {\it shear}, denoted by parameters
$\gamma_{1}$ and $\gamma_{2}\,$. This observable distortion depends upon, and
can be used to map, the distribution of baryonic and dark matter in the
Universe \citep[for a review, see][]{dmrev}.

Dedicated `Stage IV' weak-lensing surveys such as the LSST\footnote{Vera C.\
Rubin Observatory's Legacy Survey of Space and Time:
\url{http://www.lsst.org/}} \citep{LSSTOverviwe2019}, Euclid\footnote{ Euclid
satellite mission: \url{http://sci.esa.int/euclid/}} \citep{Euclid2011}, and
NGRST\footnote{ Nancy Grace Roman Space Telescope:
\url{http://roman.gsfc.nasa.gov/}} \citep{WFIRST15} are being deisgned to
constrain cosmology with unprecedented precision. To ensure that systematic
biases are within statistical uncertainty, they require estimators of the
applied shear
\begin{equation} \label{eq_shear_biases}
    \hat{\gamma}_{1,2}=(1+m)\,\gamma_{1,2}+c_{1,2}
\end{equation}
to be measured from the noisy image of each galaxy, with multiplicative bias
$\abs{m}\lesssim0.003$\, and additive bias $c_{1,2}\lesssim10^{-4}$
\citep{cropper13,euclidrequirement1,LSSTRequirement2018}. In addition, these
surveys will take a large amount of data, e.g.\ the LSST will produce about
$20$ terabytes (TB) of raw data per night. Here we develop a shear estimator
that is both accurate and fast enough to process the data from these surveys.

A practical shear measurement method must overcome many challenges to work on
real astronomical imaging: including the detection and selection of distant
galaxies, correction of detector effects and the wavelength-dependent
point-spread function (PSF), and the combination of noisy pixellated data
\citep[for a review, see][]{revRachel17}. Some methods
\citep{HSC1-GREAT3Sim,KiDS-imgSim19,HSC3-catalog,hoekstra21} advocate empirical
calibration of shear estimators such as \reGauss{} \citep{Regaussianization}
and \lensfit{} \citep{lensfit1,lensfit2} using simulated images. That merely
shifts the challenge to one of accurate simulation. The various processes
interact in complex ways, and it can be difficult to disentangle their effects.

In this paper, we extend the Fourier Power Function Shapelets
\citep[\FPFS{};][]{FPFS-Li2018} method for shear measurement. We use analytic
features of the method to eliminate bias for isolated well-sampled images of
isolated galaxies, when the PSF is accurately known and independent of
wavelength. This is only one more step towards a full pipeline, but the tests
in this paper already separate out and require control of
\vspace{-2mm}
\begin{itemize}
    \item `model bias' \citep{modelBias-Bernstein10}, due to incorrect
        assumptions about galaxy morphology,
    \item `noise bias' \citep{noiseBiasRefregier2012}, where noise terms in a
        non-linear expression for $\hat{\gamma}$ are now removed up to second
        order, and
    \item `selection bias' \citep{k2k} due to the detection of galaxies based
        on their appearance \textit{after} both a shear and instrument PSF have
        been applied.
\end{itemize}
The \FPFS{} shear estimator uses four shapelet modes
\citep{shapeletsI-Refregier2003} of a galaxy's Fourier power function
\citep{Z08} to estimate shear distortion. It avoids the need to artificially
truncate the shaplet series and thus is free from `model bias'
\citep{FPFS-Li2018}. A weighting parameter is introduced so that the `noise
bias' is proportional to the inverse square of this free parameter for faint
galaxies, and `noise bias' can be controlled by tuning the weighting parameter.
In this work, we further correct the second-order `noise bias' in the \FPFS{}
shear estimator and remove the `selection bias'.
\xlrv{
In addition, we show that \FPFS{} shear estimator is robust to stellar
contamination in the galaxy sample if the PSF is well determined.
}

Several other existing methods also meet the Stage IV weak-lensing surveys'
requirement on systematics control for isolated galaxies, including \metacal{}
\citep{metacal-Huff2017,metacal-Sheldon2017}, \FQ{} \citep{Z17,LZ2021}, \BFD{}
\citep{BFD-Bernstein2014, BFD-Bernstein2016} without relying on calibration
using external image simulations. Continuing to develop several methods
simultaneously remains a useful mitigation of risk. Among these, \FPFS{}
remains promising because it requires no prior information about galaxy
morphologies, is more than $100$ times faster than \metacal{}, processing over
a thousand galaxy images per CPU second.

This paper is organized as follows: In Section~\ref{sec:Method}, we review the
previous formalism of the \FPFS{} method, and analytically derive the
second-order correction for nonlinear noise bias and the correction for
selection bias. In Section~\ref{sec:sim}, we introduce simulated images of
galaxies used to verify (not calibrate) the performance of the proposed shear
estimator on isolated galaxies. In Section~\ref{sec:res}, we present the
results of the tests. In Section~\ref{sec:Summary}, we summarize and discuss
future work.

\section{METHOD}
\label{sec:Method}
\input{method}

\section{IMAGE SIMULATION}
\label{sec:sim}
\input{simulation}

\section{RESULTS}
\label{sec:res}
\input{results}

\section{SUMMARY AND OUTLOOK}
\label{sec:Summary}

In this paper, we improve the \FPFS{} weak lensing shear estimator, by
implementing corrections for two dominant biases. First, with an assumption
that noise in an image is a homogeneous Gaussian random field, we correct for
shear measurement noise bias to second order. Second, we derive analytic
expressions to remove selection biases due to both anisotropic shape noise and
anisotropic measurement error. Crucially, the analytic corrections that we
implement in \FPFS{} do not rely upon slow and computationally expensive
iterative processes, or upon calibration via external simulations. Our
publicly-available code (\url{https://github.com/mr-superonion/FPFS}) can
process more than a thousand galaxy images per CPU second.

Using mock imaging of isolated SNR$\gtrsim$10 galaxies with known shear, we
demonstrate that we have improved the method's accuracy by an order of
magnitude. \FPFS{} now meets the science requirements for a Stage~IV weak
lensing survey \citep[e.g.][]{cropper13,LSSTRequirement2018}.

Future work should revise this paper's assumption that galaxies are isolated.
\cite{FPFS-Li2018} and \cite{DESY3-BlendshearCalib-MacCrann2021} report that
the blending of light between neighboring galaxies on the projected plane
causes a few percent multiplicative bias for deep ground-based imaging surveys
e.g.\ the HSC\footnote{Hyper Suprime-Cam: \url{https://hsc.mtk.nao.ac.jp/ssp/}}
Survey \citep{HSC1-data}, DES\footnote{Dark Energy Survey:
\url{https://www.darkenergysurvey.org/}} \citep[DES;][]{DES16}, and the future
LSST. The bias from blending includes shear-dependent blending identification
\citep{metaDet-Sheldon2020} and bias related to redshift-dependent shear
distortion \citep{DESY3-BlendshearCalib-MacCrann2021}. \metadet{}
\citep{metaDet-Sheldon2020} is an improved version of \metacal{}, able to
correct for bias due to shear dependent blending identification. Correction for
biases related to blending in \FPFS{} should be the next effect to be tackled.
After that, we also intend to explore the use of shapelet modes of order $>2$,
which should contain independent information on the shear signal.

\section*{ACKNOWLEDGEMENTS}
% Entry for the table of contents, for this guide only
\addcontentsline{toc}{section}{ACKNOWLEDGEMENTS}

We thank Jun Zhang, Daniel Gruen for their helpful comments. XL thanks people
in IPMU and UTokyo --- Minxi He, Nobuhiko Katayama, Wentao Luo, Masamune Oguri,
Masahiro Takada, Naoki Yoshida, Chenghan Zha --- and people in the HSC
collaboration --- Robert Lupton, Rachel Mandelbaum, Hironao Miyatake, Surhud
More --- for valuable discussions. In addition, we thank the anonymous referee
for feedback that improved the quality of the paper.

XL was supported by the Global Science Graduate Course (GSGC) program of the
University of Tokyo and JSPS KAKENHI (JP19J22222). RM is supported by the UK
Space Agency through grant ST/W002612/1. The Flatiron Institute is supported by
the Simons Foundation.

\section*{DATA AVAILABILITY}
% Entry for the table of contents, for this guide only
\addcontentsline{toc}{section}{DATA AVAILABILITY}
The code  used for image processing and galaxy image simulation in this paper
is available from \url{https://github.com/mr-superonion/FPFS}.

\bibliographystyle{mnras}
\bibliography{citation}
\appendix
\input{appendix.tex}

\label{lastpage}
\end{document}

%% file: method.tex
The \citet{FPFS-Li2018} implementation of \FPFS{} measures shear from four
shapelet coefficients \citep{shapeletsI-Refregier2003} of each galaxy's Fourier
power function \citep{Z08}. We review that method in
Section~\ref{subsec:Estimator}. However, the noise bias is proportional to the
inverse square of a tuning parameter for faint galaxies. In addition, it
assumes galaxies selected to be in a sample have (isotropically) random
intrinsic orientations, and the measurement error from photon noise does not
prefer any direction. We correct nonlinear noise bias to second-order in
Section~\ref{subsec:noiseBias}, then correct two selection biases in
Section~\ref{subsec:selectBias}. Throughout this section, we use accents to
denote measurable quantities under different conditions and at different stages
of image processing; this notation is summarized for reference in
Table~\ref{tab:notation}. We will introduce shear reponses of observables and
selections, our notation for which is summarized for reference in
Table~\ref{tab:notation2}.

\begin{table}
\caption{
    Table for accent notations. The examples are for the ellipticity, but the
    notations also apply to other \FPFS{} quantities in Sec.~\ref{sec:Method}.
    }
\begin{center}
\begin{tabular}{cl} \hline
    Ellipticity with different accents & Definition \\ \hline
    $\bar{e}_{1,2}$         & intrinsic ellipticity of galaxies \\ \hline
    $e_{1,2}$               & ellipticity of lensed galaxies \\ \hline
    $\tilde{e}_{1,2}$       & ellipticity of noisy lensed galaxies \\ \hline
    $\hat{e}_{1,2}$         & ellipticity after noise bias correction\\ \hline
\end{tabular}
\end{center}
\label{tab:notation}
\end{table}

\begin{table*}
\caption{
    Table for notations of responses that will be introduced in
    Sec.~\ref{sec:Method}.
    }
\begin{center}
\begin{tabular}{cl} \hline
    Linear response to small distortion& Definition \\ \hline
    $R^\gamma_e$                &   shear ($\gamma$) response of \FPFS{}
                            ellipticity ($e$) at single galaxy level\\ \hline
    $\res^\gamma_e$             &   shear ($\gamma$) response of average
                            \FPFS{} ellipticity ($e$) for a galaxy population\\ \hline
    $\res^\gamma_\mathrm{sel}$  &   shear ($\gamma$) response of a selection from
                            a galaxy population\\ \hline
\end{tabular}
\end{center}
\label{tab:notation2}
\end{table*}

\subsection{Li et al.\ (2018)'s \FPFS{} shear estimator}
\label{subsec:Estimator}

Starting with a noiseless galaxy image $f_\vx$ in an $N\times N$
postage stamp\footnote{
To avoid bias, \citet{FPFS-Li2018} found that the radius of a circular postage
stamp around an isolated galaxy should be at least four times the galaxy's
radius determined by it's \reGauss{} moments \citep{Regaussianization}. In this
paper, we use a large postage stamp with $N=32$\,pixels. The optimal postage
stamp size, which adapts to the size of each galaxy and the presence of
neighbours, will be discussed in future work.
}, we first calculate its $2$D Fourier transform
\begin{equation}
f_\vk=\int f_\vx e^{-i\vk \cdot \vec{x}} d^2x,
\end{equation}
and its Fourier power function\footnote{We choose not to
normalize the power function by the area of the image as in the power spectral
density, as the information from an isolated galaxy does not scale as the image
size. We shall later adopt the same convention for the noise power for
consistency.}
\begin{equation} \label{eq:FourPowDef}
F_\vk=f_\vk f_\vk^* = \abs{f_\vk}^2,
\end{equation}
where the asterisk denotes a complex conjugate.  We focus on the well-sampled
case for ground-based telescope, in which the pixel size is less than the
Nyquist sampling rate; therefore, we can approximate signals in configuration
space as a continuous form. To compute the Fourier transform, we assume
periodic boundary conditions. Consequently, $\vk$ is enumerable, and $f_\vk$ is
in a discrete form.
Note that the Fourier power function is always Hermitian symmetric and the
centroid of the galaxy power function is always at $\vk=0$. This will eliminate
bias due to anisotropic mis-centering on the intrinsic source plane, caused by
noise or PSF.

\begin{figure}
\begin{center}
    \includegraphics[width=0.45\textwidth]{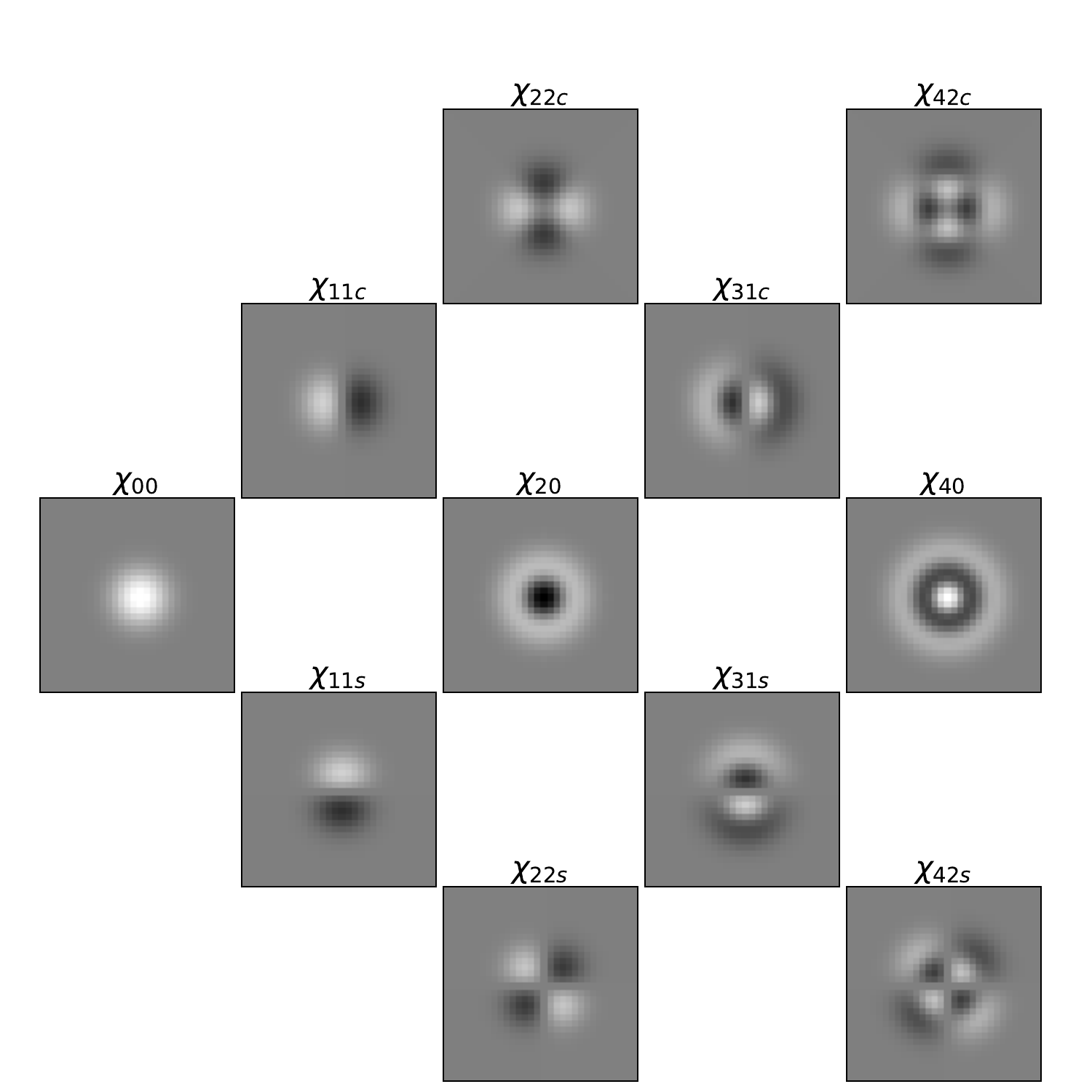}
\end{center}
\caption{
    Shapelet basis functions $\chi_{nm}$. The indices ``c'' and ``s'' refer to
    the real part and imaginary part of the complex shapelet basis functions.
    The \FPFS{} shear estimator combines projections on four shapelet basis
    functions: $\chi_{00}$, $\chi_{22c}$, $\chi_{22s}$, and $\chi_{40}$. The
    linear gray color map ranges from $-0.18$ (black) to $0.18$ (white).
    }
    \label{fig:shapeletBasis}
\end{figure}

In Fourier space, the effect of blurring by a PSF can be removed by dividing
$F_\vk$ by the PSF's Fourier power function, $G_\vk$ \citep{Z08}.  We decompose
the result into shapelets\footnote{
Since the shapelet basis vectors are orthogonal
\citep{shapeletsI-Refregier2003}, and not convolved with the PSF \citep[as
in][]{polar_Shapelets}, shapelet modes can be estimated by a direct projection
following equation~\eqref{Shapelets_decompose} --- we do {\it not} need to know
the projections on any other shapelet basis vectors to estimate one shapelet
mode, and avoid the need to artificially truncate the shapelet series.
} \citep{polar_Shapelets}
\begin{equation}\label{Shapelets_decompose}
    M_{nm}=\sum_\vk \chi_{nm}^{*}(\vk) \frac{F_\vk}{G_\vk}\,,
\end{equation}
where, to remove a negative sign later, we use complex conjugates of the polar
shapelet basis functions (Figure~\ref{fig:shapeletBasis})
\begin{equation}\label{eq:shapeletmodes}
\begin{split}
\chi_{nm}(\rho,\phi)&=\frac{(-1)^{(n-|m|)/2}}{r_F^{|m|+1}}\left\lbrace
    \frac{[(n-|m|)/2]!}{\pi[(n+|m|)/2]!}\right\rbrace^\frac{1}{2}\\
    &\times
    \rho^{|m|}L^{|m|}_{\frac{n-|m|}{2}}\left(\frac{r^2}{r_F^2}\right)e^{-\rho^2/2r_F^2}
    e^{-im\phi},
\end{split}
\end{equation}
where $L^{p}_{q}$ are the Laguerre Polynomials, $n$ is the radial number and
$m$ is the spin number, and $r_F$ determines the scale of shapelet functions in
Fourier space. In addition to $\vk$, we use polar coordinates ($\rho$, $\phi$)
to denote discrete locations in $2$D Fourier space.  The scale size of
shapelets used to represent the galaxy, $r_{F}$, should be less than the scale
radius of PSF's Fourier power, $r_\mathrm{P}$, to avoid boosting small-scale
noise during PSF deconvolution. In this paper, we fix $r_{F}/
r_\mathrm{P}=0.75$.

To first order in shear, the shear distortion operator correlates a {\it
finite} number of shapelet modes (separated by $|\Delta n|=2$ and $|\Delta
m|=2$ \citealt{polar_Shapelets}).  It is therefore possible to estimate an
applied shear signal using a {\it finite} number of shapelet modes. The
\citet{FPFS-Li2018} \FPFS{} algorithm uses four shapelet modes to construct its
shear estimator, with \FPFS{} `\textit{ellipticity}',
\begin{align}\label{ellipticity_define}
e_1\equiv\frac{M_{22c}}{M_{00}+C},\qquad
e_2\equiv\frac{M_{22s}}{M_{00}+C}\,,
\end{align}
where $M_{22c}$ and $M_{22s}$ refer to the real (`$\cos{}$') and imaginary
(`$\sin{}$') components of the complex shapelet mode $M_{22}$ respectively. The
positive constant parameter $C$ adjusts the relative weight between galaxies of
different luminosities. A large value of $C$ puts more weight on bright
galaxies, and suppresses noise bias, while a small value of $C$ weights
galaxies more uniformly. We similarly define a few useful spin $m=0$ \FPFS{}
quantities:
\begin{equation}
    s_{0,2,4}=\frac{M_{00,20,40}}{M_{00}+C}\,.
    \label{s}
\end{equation}
In principle, the value of $C$ can be different in each of these quantities;
however, here we set them all to $2.5\,\sigma_{M_{00}}$ for simplicity, where
$\sigma_{M_{00}}$ is the standard deviation of measurement error on $M_{00}$
caused by photon noise on galaxy images. The details of tuning the weighting
parameter is shown in Sections~\ref{subsec:res_noisebias} and
\ref{subsec:res_error}. The `\textit{flux ratio}' $s_0$ was suggested by
\citet{FPFS-Li2018} as a criterion to select a galaxy sample, helping to remove
faint galaxies and spurious detections, etc.

When the image of a galaxy is distorted by shear $\gamma_\beta$, its
ellipticity transforms to first order in $\gamma$ as
\begin{equation}\label{ellipticity_transform}
    \bar{e}_\alpha\rightarrow{e}_\alpha=
    \bar{e}_\alpha+ \sum_{\beta=1,2} \gamma_\beta(R^\gamma_e)_{\alpha\beta},
\end{equation}
where components $\alpha\in\{1,2\}$ and $\beta\in\{1,2\}$, and the
`\textit{shear responsivity}' $(R^\gamma_e)_{\alpha\beta}=\partial
e_\alpha/\partial\gamma_\beta\,$ \citep{metacal-Sheldon2017}. The shear
responsivity of the \FPFS{} ellipticity for an individual galaxy is a scalar
quantity\footnote{Equation~\eqref{response_define} differs by a minus sign from
that in \citet{FPFS-Li2018}, who defined the shear response with respect to
the shear distortion in Fourier space.  All the shear responses in this
paper are defined with respect to the shear distortion in configuration
space.},
\begin{align}\label{response_define}
R^\gamma_e=\frac{1}{\sqrt{2}}\left(s_4-s_0-e_{1}^2-e_{2}^2\right),
\end{align}
plus terms involving spin $m=4$ modes --- which all average to zero when we now
average $R^\gamma_e$ over all galaxies in a sample, to obtain a population
responsivity (denoted with curly letters) $\res^\gamma_e\equiv\langle
R^\gamma_e \rangle$. We intend to investigate in future work
$\mathcal{O}(\gamma^2)$ terms in $\res^\gamma_e$ that may not necessarily be
small  near galaxy clusters. Since the difference between $\langle \bar{e}_1^2
\rangle$ and $\langle \bar{e}_2^2 \rangle$ is negligible, we define the shear
response as the average responses of two shear components and do not
distinguish between the responses of two shear components.

Assuming that the galaxy population is selected such that
$\langle \bar{e}_\alpha \rangle=0$ in the absence of shear,
equation~\eqref{ellipticity_transform} suggests a shear estimator
\citep{FPFS-Li2018}
\begin{equation}\label{eq:estimator_0}
\hat{\gamma}_{\alpha}\equiv\frac{\left\langle e_{\alpha}
\right\rangle} {\res^\gamma_e}
    =\cancelto{0}{\frac{\left\langle \bar{e}_{\alpha} \right\rangle}
    {\res^\gamma_e}} + \gamma_\alpha
    =\gamma_\alpha\,.
\end{equation}
The population variance of $\bar{e}_\alpha$ is known as the intrinsic shape
noise.

\subsection{Nonlinear noise bias}
\label{subsec:noiseBias}

\subsubsection{Effect of observational noise}

In real observations the galaxy image is contaminated by photon noise and read
noise. We denote the total image noise in Fourier space as $n_\vk$, i.e.\
the {observed} galaxy image is
\begin{equation}
f^\mathrm{o}_\vk = f_\vk + n_\vk,
\end{equation}
and the observed galaxy Fourier power function is
\begin{equation}
F^\mathrm{o}_\vk = f^\mathrm{o}_\vk (f^\mathrm{o}_\vk)^* = \abs{f^\mathrm{o}_\vk}^2.
\end{equation}

We shall assume that the source and image noise do not correlate, i.e.\
$\langle f_\vk n_\vk \rangle=0\,$.  We shall also assume that the noise is
dominated by background and read noise, so that we can neglect photon noise on
the galaxy flux\footnote{\cite{LZ2021} present a formalism to include galaxy
photon noise, but we proceed here on the assumption that it is negligible for
the faint galaxies that are most affected by noise bias, and verify the
validity of this in Section~\ref{subsec:res_selbias}.}. This makes the noise a
mean zero homogeneous Gaussian random field that, averaged over different noise
realizations, has Fourier power function with expectation value
\begin{equation} \label{eq_noise_power}
\mathcal{N}_\vk
%\equiv\left\langle{N}_\vk\right\rangle
    =\left\langle{n}_\vk{n}_\vk^{*}\right\rangle\,.
\end{equation}
Note that this is different from $F_\vk$ which is defined for each galaxy,
and we differentiate them with different fonts.
Using Isserlis' theorem (also known as Wick's theorem in quantum field theory),
its $4$-point correlation function is
\begin{equation}\label{eq_IsserlisTheorem}
    \left\langle{n}_\vk{n}_{\vk'}{n}_\vk^*{n}_{\vk'}^*\right\rangle
    =\left(1+\deltaK(\vk-\vk')+\deltaK(\vk+\vk')\right)
    \mathcal{N}_\vk \mathcal{N}_{\vk'}\,,
\end{equation}
where $\deltaK$ denotes the Kronecker delta function.

To account for image noise \citep[following][]{Z15}, the expectation value of
the noise power \eqref{eq_noise_power} can be measured from blank patches of
sky, and subtracted from an observed galaxy's Fourier power function.  Using a
tilde to label corrected quantities, this yields\!\!
\begin{equation}
    \tilde{F}_\vk=F^o_\vk-\mathcal{N}_\vk\,,
\end{equation}
from which $\tilde{e}_\alpha$, $\tilde{s}_0$, etc.\ can be defined similarly
as in \eqref{ellipticity_define}, \eqref{s}, and \eqref{Shapelets_decompose}.
Note the accent notation of \FPFS{} flux ratio, $s$, follows that of \FPFS{}
ellipticity, $e_{1,2}$, shown in Table~\ref{tab:notation}. However, compared to
the noiseless image power function $F_\vk$, this now contains {residual noise
power}
\begin{equation}\label{eq_ps_res_gal}
    {\epsilon}_\vk \equiv \tilde{F}_\vk - F_\vk
    ={n}_\vk{n}_\vk^*-\mathcal{N}_\vk
    +f_\vk n_\vk^{*}+f_\vk^{*} n_\vk.
\end{equation}
Across a galaxy population, the expectation value of residual noise is zero, $
\langle \epsilon_\vk \rangle=0$. For any individual galaxy, the residual noise
power includes contributions from the individual realisation of noise, and any
correlation between that noise and the galaxy flux \citep[see also][]{LZ2021}.
Combining equations~\eqref{eq_IsserlisTheorem} and \eqref{eq_ps_res_gal}, the
two-point correlation function of the residual noise power is
\begin{equation}\label{eq_var_res_noisyGal}
\left\langle \epsilon_\vk\epsilon^*_{\vk'} \right\rangle=
    \left(\deltaK(\vk-\vk')+\deltaK(\vk+\vk')\right)
    (\mathcal{N}_\vk^2+2F_\vk \mathcal{N}_\vk)\,.
\end{equation}

During shape measurement, the galaxy power function (which now includes
residual noise) is divided by the PSF power and decomposed into shapelets
(equation~\ref{Shapelets_decompose}). The shapelet modes of the residual noise
are
\begin{equation}
    \mathcal{E}_{nm}=\sum_\vk \chi_{nm}^{*}(\vk) \frac{\epsilon_\vk}{G_\vk}.
\end{equation}
Again, their expectation values $\langle \mathcal{E}_{nm} \rangle$ vanish
because $\langle \epsilon_\vk \rangle=0\,$. For an individual galaxy however,
the their covariance is
\begin{equation}\label{eq:shapeletcov}
\begin{split}
\mathcal{V}_{nmn'm'}
&\equiv \bigl\langle \mathcal{E}_{nm} \mathcal{E}^*_{n'm'} \bigr\rangle \\
&= \sum_\vk \left(\frac{\chi^*_{nm}\chi_{n'm'}+\chi_{nm}\chi_{n'm'}}
    {G^2_\vk}\right)
    (\mathcal{N}_\vk^2+2F_\vk \mathcal{N}_\vk)\,.
\end{split}
\end{equation}
The shapelet modes thus become correlated ($V_{nmn'm'} \neq 0$) due to
inhomogeneous and anisotropic residual noise, and anisotropy in the PSF.
The covariances can be measured from nearby blank patches of sky, the
galaxy itself, and the PSF model.

\subsubsection{Correction for noise bias}

Propagating the contribution of residual noise into \FPFS{} ellipticity
estimators yields an expectation values
\begin{eqnarray}
\left\langle \tilde{e}_1 \right\rangle \equiv \left\langle\frac{M_{22c}+\mathcal{E}_{22c}}
{M_{00}+C+\mathcal{E}_{00}}\right\rangle, &&
\left\langle \tilde{e}_2 \right\rangle \equiv \left\langle\frac{M_{22s}+\mathcal{E}_{22s}}
{M_{00}+C+\mathcal{E}_{00}}\right\rangle\,.~~~~~~~~
\end{eqnarray}
Expanding the \FPFS{} ellipticity as Taylor series of
$\frac{\mathcal{E}_{00}}{M_{00}+C}$ about the point $\mathcal{E}_{00}=0$ and
inserting the covariance of shapelet modes (equation~\eqref{eq:shapeletcov}),
this is
\begin{equation}
    \label{eq:noiseCorrE_0}
    \begin{split}
    \left\langle \tilde{e}_1\right\rangle
    &=\left\langle e_1\left(1+
        \frac{\mathcal{V}_{0000}}
        {(M_{00}+C)^2}\right)
    -\frac{\mathcal{V}_{0022c}}
    {(M_{00}+C)^2}
    +O\left(\left(\frac{\mathcal{E}_{nm}}{M_{00}+C}\right)^4\right)
    \right\rangle,\\
    \left\langle \tilde{e}_2\right\rangle
    &=\left\langle e_2\left(1+
        \frac{\mathcal{V}_{0000}}
        {(M_{00}+C)^2}\right)
    -\frac{\mathcal{V}_{0022s}}
    {(M_{00}+C)^2}
    +O\left(\left(\frac{\mathcal{E}_{nm}}{M_{00}+C}\right)^4\right)
    \right\rangle,
    \end{split}
\end{equation}
where $\mathcal{V}_{0022c}$ ($\mathcal{V}_{0022s}$) refers to the covariance
between $\mathcal{E}_{00}$ and $\mathcal{E}_{22c}$ ($\mathcal{E}_{22s}$). One
can notice immediately that the second-order terms are in the form of additive
and multiplicative biases, proportional to the inverse square of $M_{00}+C\,$.
And we shall neglect terms of fourth-order and higher.

Therefore one version of the \FPFS{} ellipticity suitably corrected for noise
bias up to second-order is
\begin{equation}
    \label{eq:noiseCorrE}
    \begin{split}
    \hat{e}_1
    &= \frac{1}{T} \left(\tilde{e}_1+\frac{\mathcal{V}_{0022c}}
    {(\tilde{M}_{00}+C)^2}\right) ,\\
    \hat{e}_2
    &= \frac{1}{T} \left(\tilde{e}_1+\frac{\mathcal{V}_{0022s}}
    {(\tilde{M}_{00}+C)^2}\right) ,
    \end{split}
\end{equation}
where $T=1+ \frac{\mathcal{V}_{0000}} {(\tilde{M}_{00}+C)^2}\,$. Similarly,
the \FPFS{} flux ratio measured from a noisy image, $\tilde{s}_0$, is also
subject to noise bias, but can be corrected as
\begin{equation}
    \hat{s}_0
    = \frac{1}{T} \left(\tilde{s}_0 +\frac{\mathcal{V}_{0000}}
    {(\tilde{M}_{00}+C)^2}\right) .
\end{equation}
The corrected shear responsivity for a single, noisy galaxy becomes
\begin{align}\label{response_corrected}
\hat{R}^\gamma_e=\frac{1}{\sqrt{2}}\left(\hat{s}_4-\hat{s}_0-\hat{e}_{1}^2-\hat{e}_{2}^2\right),
\end{align}
where corrected versions of the other \FPFS{} quantities are listed in
Appendix~\ref{app:noiseBias}.

Once again, we average responsivity across a galaxy sample to obtain population
response $\hat{\res}^\gamma_e=\langle\hat{R}^\gamma_e\rangle$.  Assuming
that $\langle\bar{e}_\alpha\rangle=0$, and now also that
$\langle\delta{e}_\alpha\rangle=0$, where
\begin{align}\label{eq:deltaElli}
    \delta{e}_{\alpha} \equiv \hat{e}_{\alpha}-e_{\alpha}
\end{align}
is measurement error due to image noise, we obtain a new shear estimator
\begin{equation}\label{eq:estimator_A}
    \hat{\gamma}^{\mathrm{A}}_{\alpha}
    \equiv\frac{\left\langle \hat{e}_{\alpha}
    \right\rangle} {\hat{\res}^\gamma_e}
    =\cancelto{0}{\frac{\left\langle \bar{e}_{\alpha} \right\rangle}
    {\hat{\res}^\gamma_e}}
    +\cancelto{0}{\frac{\left\langle \delta{e}_{\alpha} \right\rangle}
    {\hat{\res}^\gamma_e}}
    +\gamma_\alpha
    =\gamma_\alpha
\end{equation}
that corrects for nonlinear noise bias to second order.  We label this first
estimator `A' because we shall next propose more complications. The performance
of this shear estimator will be tested in Section~\ref{subsec:res_noisebias}.

\subsection{Selection bias}
\label{subsec:selectBias}

For a complete sample of galaxies, the assumptions that
$\langle\bar{e}_\alpha\rangle=0$ and $\langle\delta{e}_\alpha\rangle=0$ are
statistically correct. For an incomplete sample, perhaps restricted to
galaxies above a threshold in \FPFS{} flux ratio $\hat{s}_0$, this assumption
can be broken --- creating a shear estimation bias known as selection bias. We
correct for the selection bias due to anisotropic intrinsic shape noise,
$\langle \bar{e}_{\alpha} \rangle\neq0$, in Section~\ref{subsec:selBias_bar},
and for the seelction bias due to anisotropic measurement error, $\langle
\delta{e}_{\alpha} \rangle\neq0$, in Section~\ref{subsec:selBias_check}.

\subsubsection{Selection bias due to anisotropic intrinsic shape noise}
\label{subsec:selBias_bar}

Here we derive a correction for the selection bias that is introduced if
$\langle \bar{e}_{\alpha} \rangle\neq0$ in a galaxy population. This can occur
if the population is selected according to some quantity that is changed by
shear. For example, although unbiased selection could be based upon galaxies'
intrinsic \FPFS{} flux ratios $\bar{s}_0$, it would in practice be based upon
their lensed flux ratios $s_0$. To first order in $\gamma$, these transform
under shear as
\begin{equation} \label{s0_transform}
    \bar{s}_0\rightarrow s_0=\bar{s}_0+ \sum_{\alpha=1,2} \left.
    \frac{\partial s_0}{\partial \gamma_\alpha}
    \right|_{s_0=\bar{s}_0} \gamma_\alpha.
\end{equation}
The difference may cause individual galaxies to cross selection thresholds, or
to change weight. This effect was first studied in \citet{k2k} and is referred
to as Kaiser flow.  In the following discussion, we shall neglect terms
$\mathcal{O}(\gamma^2)$, and temporarily also neglect shape measurement noise.

First consider a complete population of galaxies, whose intrinsic \FPFS{} flux
ratio, $\bar{s}_0$, and intrinsic ellipticity, $\bar{e}_\alpha$ are distributed
with probability density function (PDF) $\bar{P}(\bar{s}_0, \bar{e}_\alpha)$.
Because there is no preferred direction in the Universe, the expectation value
of intrinsic galaxy ellipticity $\langle \bar{e}_\alpha\rangle=0$.

Next let us identify a subset of the population.  In the absence of shear, this
can be selected via a cut ${s}_0^{\mathrm{low}}< \bar{s}_0<
{s}_0^{\mathrm{upp}}$ between lower and upper bounds on the {intrinsic} \FPFS{}
flux ratio.  Because the selection criterion is a spin-$0$ quantity without
preferred direction, the expectation value of intrinsic galaxy ellipticity must
be preserved\!\!
\begin{equation}
    \left\langle \bar{e}_\alpha\right\rangle\Big|_{\gamma_\alpha=0}
    =\iint_{s_0^{\mathrm{low}}}^{s_0^{\mathrm{upp}}}
    \bar{P}\left( \bar{s},\bar{e}_\alpha \right) \bar{e}_\alpha ~\rmd\bar{s}
    \rmd\bar{e}_\alpha=0\,.
\end{equation}
However, if a shear has been applied, the selection $\bar{s}_0^{\mathrm{low}}<
{s}_0< \bar{s}_0^{\mathrm{upp}}$ must be between limits on {\it lensed}
quantities
\begin{equation}
    \left\langle \bar{e}_\alpha\right\rangle
    =\iint_{s_0^{\mathrm{low}}}^{s_0^{\mathrm{upp}}}
    \bar{P}\left( \bar{s},\bar{e}_\alpha \right) \bar{e}_\alpha ~\rmd{s}
    \rmd\bar{e}_\alpha\,.
\end{equation}
Using equation~\eqref{s0_transform}, this selection is equivalent to modified
bounds on the {\it intrinsic} source plane
\begin{equation} \label{ugly_limits}
    \left\langle \bar{e}_\alpha\right\rangle
    =\iint_{s_0^\mathrm{low}-\left.\frac{\partial s_0}{\partial\gamma_{\alpha}}
    \right|_{s_0=s_0^\mathrm{low}} \gamma_\alpha}^{s_0^\mathrm{upp}-\left.\frac{\partial s_0}{\partial\gamma_{\alpha}}
    \right|_{s_0=s_0^\mathrm{upp}} \gamma_\alpha}
    \bar{P}\left( \bar{s},\bar{e}_\alpha \right) \bar{e}_\alpha ~\rmd\bar{s}
    \rmd\bar{e}_\alpha \,,
\end{equation}
which can have spurious anisotropy $\langle \bar{e}_\alpha\rangle\neq0$. We
define the shear response of the galaxy selection on the galaxy sample level,
$\res^\gamma_\mathrm{sel}$, as the ratio between this anisotropy versus the
shear that caused it:
\begin{equation}\label{eq:select_responsivity}
\begin{split}
    \res^\gamma_\mathrm{sel}&\equiv\frac{\langle \bar{e}_\alpha \rangle-\langle \bar{e}_\alpha \rangle|_{\gamma_\alpha=0}}
    {\gamma_\alpha}=\frac{\langle \bar{e}_\alpha \rangle}{\gamma_\alpha} \\
    &=-\bar{P}(s_0^\mathrm{upp}) \left.\left\langle \bar{e}_\alpha
    \frac{\partial s_0}{\partial \gamma_\alpha}\right\rangle
    \right|_{s_0=s_0^\mathrm{upp}}
    +\bar{P}(s_0^\mathrm{low}) \left.\left\langle \bar{e}_\alpha
    \frac{\partial s_0}{\partial \gamma_\alpha}\right\rangle
    \right|_{s_0=s_0^\mathrm{low}}\,,
\end{split}
\end{equation}
where $\bar{P}(s_0^\mathrm{upp})$ and $\bar{P}(s_0^\mathrm{low})$ are the
marginal probability distributions of $s_0$ at $s_0^\mathrm{upp}$ and
$s_0^\mathrm{low}$, respectively. Those marginal probability distributions can
be estimated approximately by the average marginal probability distributions in
$[\,s_0^\mathrm{upp}-0.01\,, s_0^\mathrm{upp}+0.01\,]$ and
$[\,s_0^\mathrm{low}-0.01\,, s_0^\mathrm{low}+0.01\,]\,$, respectively. Note
that in real observations, we are only able to estimate the marginal
probability distributions from noisy, lensed galaxies instead of from
noiseless, intrinsic galaxies. We take the assumption that the difference
between the intrinsic, noiseless marginal probability distributions and lensed,
noisy probability distributions is negligible. Using equation~(18) of
\citet{FPFS-Li2018},
\begin{equation}\label{eq:response_s0}
\left. \left\langle \bar{e}_\alpha \frac{\partial s_0}
    {\partial \gamma_\alpha}\right\rangle \right |_{s_0=s^\mathrm{low}_0}
    =\sqrt{2}\left. \left\langle (e_\alpha^2)(1-s_0) \right\rangle  \right |_{s_0=s^\mathrm{low}_0}\,,
\end{equation}
we can obtain the shear response of the selection from measurable quantities.

A shear estimator incorporating this selection responsivity
\begin{equation}\label{eq:estimator_B}
    \hat{\gamma}^{\mathrm{B}}_{\alpha}
    \equiv\frac{\left\langle \hat{e}_{\alpha}
    \right\rangle} {\hat{\res}^\gamma_e+\res^\gamma_\mathrm{sel}}
    =\cancelto{0}{\frac{\left\langle \delta{e}_{\alpha} \right\rangle}
    {\hat{\res}^\gamma_e+\res^\gamma_\mathrm{sel}}}
    +\frac{\left\langle \bar{e}_{\alpha} \right\rangle}
    {\hat{\res}^\gamma_e+\res^\gamma_\mathrm{sel}}
    +\frac{\hat{\res}^\gamma_e~\gamma_\alpha}{\hat{\res}^\gamma_e+\res^\gamma_\mathrm{sel}}
    =\gamma_\alpha
\end{equation}
should then be immune to selection bias due to anisotropic intrinsic shape
noise. Its performance will be tested on galaxy image simulations in
Section~\ref{subsec:res_selbias}.

\subsubsection{Selection bias due to anisotropic measurement error}
\label{subsec:selBias_check}

Here we re-introduce image noise, and derive a correction for the selection
bias that is introduced if $\langle \delta{e}_{\alpha} \rangle\neq0$ in a
galaxy population.  This can occur if image noise leads to measurement error in
a galaxy's \FPFS{} ellipticity $\delta e_\alpha$ that correlates with
measurement error in a quantity used for sample selection, e.g.\ the \FPFS{}
flux ratio $\hat{s}_0$.

Consider first a population of galaxies with a PDF of noiseless \textit{but now
lensed} quantities $P(e_\alpha, s_0)\,$. In this lensed plane, measurement
error on the \FPFS{} flux ratio is (c.f.\ equation~\eqref{eq:deltaElli})
\begin{equation}\label{eq:deltaS0}
    \delta{s}_{0}\equiv\hat{s}_{0}-s_0\,.
\end{equation}
The definitions of $s_0$ and $e_\alpha$ ensure that, if the image noise is
Gaussian, the noise on $s_0$ and $e_\alpha$ will both be close to Gaussian. In
this case, the contribution of image noise to the population variances is
\begin{equation} \label{eq:sigmaE}
\begin{split}
    \sigma_{s}^2&=\langle (\delta{s}_0)^2 \rangle\,,\\
    \sigma_{e_\alpha}^2&=\langle (\delta{e}_\alpha)^2 \rangle\,,
\end{split}
\end{equation}
both of which can be estimated from observed galaxy images \xlrv{(see
equations~\eqref{eq:app_dede})} using the covariance of measurement errors on
shapelet modes (c.f.\ equation~\eqref{eq:shapeletcov}). The accuracy of the
variance estimate will be tested in Section~\ref{subsec:res_error}.
Furthermore, the correlation between the measurement errors is
\begin{equation} \label{eq:sigmaES}
    \rho_{e_\alpha s}=\frac{\langle \delta{e}_\alpha \delta{s}_0 \rangle}
    {\sigma_{e_\alpha} \sigma_{s}},
\end{equation}
which can also be estimated from noisy galaxy images (see
equations~\eqref{eq:app_deds}). As indicated by equations~\eqref{eq:app_deds}
and \eqref{eq:shapeletcov}, $\rho_{es}\neq0$ in this lensed plane if either the
PSF or the noise power function is anisotropic.

The PDF of {\it noisy} lensed quantities can thus be approximated by
\begin{equation}
    %\hat{P}({e}_\alpha, {s}_0) \simeq P(e_\alpha, s_0)
    %\circledast P^\delta(e_\alpha, s_0)
    \hat{P} \simeq P \circledast P^\delta
\end{equation}
where $\circledast$ refers to the convolution operation and
%\rjm{
%our assumption of symmetric Gaussian noise ensures that
%$\hat{P}(\hat{e}_\alpha, \hat{s}_0)\simeq\hat{P}({e}_\alpha, {s}_0)$
%and ${P}(\hat{e}_\alpha, \hat{s}_0)\simeq{P}({e}_\alpha, {s}_0)$ on
%average,
%}
\begin{equation}
    P^{\delta}(e_\alpha,s_0)\equiv
    \frac{1}{2\,\pi \sigma_s \sigma_{e_\alpha}}\exp\left(
        -\frac{e_\alpha^2}{2\,\sigma^2_{e_\alpha}}
        -\frac{s_0^2}{2\,\sigma^2_{s}}
        +\rho_{e_\alpha s} \frac{e_\alpha
        s_0}{\sigma_{e_\alpha} \sigma_{s}}\right)\,.
\end{equation}
Convolution with a symmetric kernel $P^{\delta}$ does not shift the centroid of
${P}$, so the average ellipticity of a complete population remains unbiased.
However, a cut on \FPFS{} flux ratio $\hat{s}_0$ in the presence of noise
\textit{can} bias the average ellipticity such that $\langle \delta{e}_{\alpha}
\rangle\neq0$, if the measurement errors $\delta{e}_\alpha$ and $\delta{s}_0$
are correlated. The resulting change in ellipticity is
\begin{equation}
    e^\Delta _{\alpha} \equiv
    \iint_{s_0^{\mathrm{low}}}^{s_0^{\mathrm{upp}}} e_\alpha
    \left( \hat{P} - P \right)
    \rmd e_\alpha \rmd s_0\,,
\end{equation}
which we propose to estimate from measurable quantities
\begin{equation}\label{eq:estimator_C_correctTermApp}
    \hat{e}^\Delta_\alpha \simeq
    \iint_{s_0^{\mathrm{low}}}^{s_0^{\mathrm{upp}}} e_\alpha
    \left( \hat{P} \circledast P^\delta - \hat{P} \right)
    \rmd e_\alpha \rmd s_0\,.
\end{equation}
For simplicity, we do not recover $P$ by deconvolving $\hat{P}\,$; instead, we
take the approximation --- $\hat{P} \simeq P\,$.

A final shear estimator correcting for noise bias and both types of selection
bias is thus
\begin{equation}\label{eq:estimator_C}
    \hat{\gamma}^\mathrm{C}_{\alpha}=
    \frac{\left\langle \hat{e}_{\alpha} \right\rangle-
    \hat{e}^\Delta_{\alpha} }
    {\hat{\res}^\gamma_e+\res^\gamma_s}=\gamma_\alpha\,.
\end{equation}
The performance of all shear estimators will be tested in
Section~\ref{subsec:res_selbias}.

%% file: simulation.tex
\subsection{Galaxies, PSF and noise}

We test the \FPFS{} shear estimators by running them on mock astronomical
images that have been sheared by a known amount. Our mock data are very similar
to sample~2 of \citet{HSC1-GREAT3Sim}, replicating the image quality and
observing conditions of the Hyper-Suprime Cam (HSC) survey on the 8\,m
ground-based Subaru telescope, whose deep coadded $i$-band images of the
extragalactic sky resolve $\sim$$20$ galaxies per arcminute$^2$ brighter than
$i=24.5$\footnote{
A cut at $i<24.5$ is applied to the real HSC shear catalog, to remove faint
galaxies and false detections. For the simulations in this paper, we force
a measurement for each input galaxy and do {\it not} apply a magnitude cut.
} \citep{HSC1-catalog,HSC3-catalog}. The pixel scale is $0\farcs168\,$.

\begin{figure}
\begin{center}
    \includegraphics[width=0.45\textwidth]{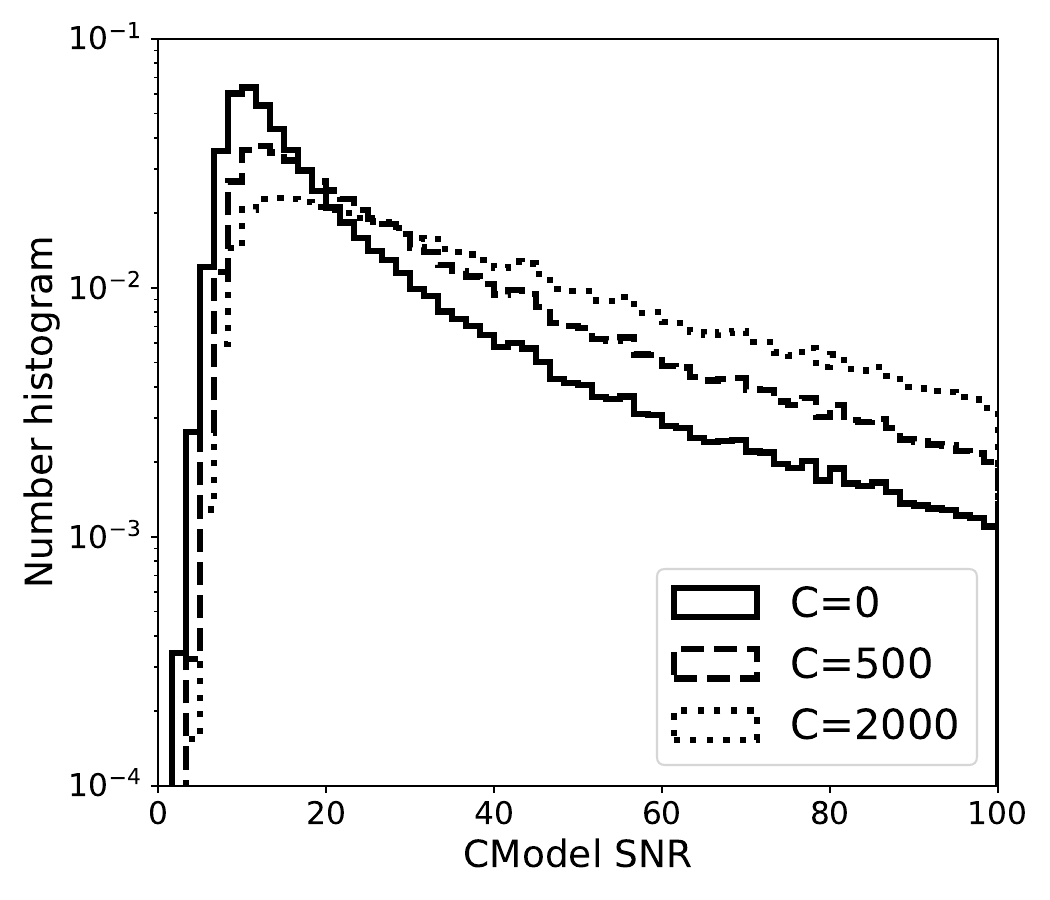}
\end{center}
\caption{
    The normalized number histogram as a function of CModel SNR. The histogram
    is weighted by \FPFS{} ellipticity weight with three different setups,
    i.e.\ $C=0$ (solid line), $C=500$ (dashed line) and $C=2000$ (dotted
    line).
    }
    \label{fig:histSNR}
\end{figure}

Galaxy images are generated using the open-source package Galsim
\citep{GalSim}. We randomly select $8\times10^4$ galaxies without repetition
from the COSMOS HST Survey
catalogue\footnote{\url{https://zenodo.org/record/3242143\#.YPBGdfaRUQV}}
\citep{HST-shapeCatalog-Alexie2007}, which has limiting magnitude $F814W=25.2$.
All galaxies have known photometric reshifts. The galaxy shapes are
approximated with the best-fitting parametric (\citealt{deVauProfile1948} or
\citealt{Sersic1963}) profile, sheared, convolved with a model of the HSC PSF,
then rendered in $64\times 64$\,pixel postage stamp images (including a border
around the $32\times 32$\,pixel region used for shear measurement). The pixel
values are finally are multiplied by $2.587$ to rescale the units, so their
$i=27$ photometric zeropoint matches that of real HSC pipeline data
\citep{HSC3-catalog}.

The image PSF is modelled as a \cite{Moffat1969} profile,
\begin{equation}\label{Moffat PSF}
    g_{m}(\mathbf{x})=[1+c(|\mathbf{x}|/r_\mathrm{P})^2]^{-3.5},
\end{equation}
where $c=2^{0.4}-1$ is a constant parameter and $r_\mathrm{P}$ is adjusted
such that the Full Width Half Maximum (FWHM) of the PSF is $0\farcs6$,
matching the mean seeing of the HSC survey \citep{HSC3-catalog}. The profile
is truncated at a radius four times larger than the FWHM. The PSF is then
sheared so that it has ellipticity $(e_1=0.02, e_2=-0.02)\,$.

We add image noise from a constant sky background and read noise. This includes
anisotropic (square-like) correlation between adjacent pixels matching the
autocorrelation function of a third-order Lanczos kernel, i.e.\ $a=3$ in
\begin{equation}
    L(x,y)=\begin{cases}
        \sinc{x/a}\,\sinc{x}\,\sinc{y/a}\,\sinc{y}
        & \mathrm{if}~\abs{x},\abs{y}<a\\
        0
        & \mathrm{otherwise,}
    \end{cases}
\end{equation}
where $\sinc{x}=\sin{(\pi x)}/\pi x$. This kernel was used to warp and co-add
images taken during the first-year HSC survey \citep{HSC1-pipeline}. Ignoring
pixel-to-pixel correlations, our resulting noise variance is $7\times 10^{-3}$,
which is about two times of the average noise variance in HSC data shown in
\citep{HSC3-catalog}. In this paper, we do not include photon noise on the
galaxy fluxes. This is to increase efficiency because the same realisation of
noise can be used in multiple images (see below). However, it means that our
tests on the effectiveness of correction for selection bias are an optimistic
limit. In Section~\ref{subsec:res_selbias} we present tests that bracket the
performance achievable if photon noise were to be included.

Our simulated images thus include galaxies with a realistic range of
signal-to-noise ratios, SNR, greater than $\sim$10 (Figure~\ref{fig:histSNR}).
We measure a galaxy's SNR using {\tt CModel} \citep{SDSSpipe2001}, which fits
each image with a linear combination of an exponential and a de Vaucouleurs
\citep{deVauProfile1948} model, as implemented in the HSC pipeline
\citep{HSC1-pipeline}.  Writing the \FPFS{} ellipticity as a weighted
dimensionless quantity $e_{1}\equiv w\,{M_{22c}}/{M_{00}}$, where
$w=(1+C/M_{00})^{-1}$, we find that a value of $C=2000$ reduces the effective
contribution of the faintest galaxies by a factor $\sim$3 (dotted line in
Figure~\ref{fig:histSNR}). However, shear measurements from faint galaxies are
noisier, and we shall find in Section~\ref{subsec:res_error} that this
weighting also optimises overall SNR.

\subsection{Shape noise cancellation}

To efficiently reduce intrinsic shape noise in our shear measurements, we
generate images of each galaxy in pairs \citep[following][]{galsim-STEP2},
where the intrinsic ellipticity of one is rotated by $90\deg$ (flipping its
sign) before applying shear. We then generate three images of each pair with
three different shears: ($\gamma_1=0.02$, $\gamma_2=0$), ($\gamma_1=-0.02$,
$\gamma_2=0$) and ($\gamma_1=0$, $\gamma_2=0$), but all with exactly the same
realisation of image noise
\citep[following][]{preciseSim-Pujol2019,metaDet-Sheldon2020}. All images are
convolved with the same PSF.

To measure the shear measurement bias (equation~\eqref{eq_shear_biases}) of an
estimator $\hat{\gamma}$, we calculate
\begin{equation}
    {c_1} = \frac{\langle \hat{e}_1^0 \rangle}
    {\langle \hat{R}^{\gamma^0}_{e} \rangle}
\end{equation}
and
\begin{equation}
    {m_1} = \frac{\langle \hat{e}_1^+ - \hat{e}_1^- \rangle}
    {0.02\langle \hat{R}^{\gamma+}_{e}
    + \hat{R}^{\gamma-}_{e} \rangle}-1\,,
\end{equation}
where $\hat{e}_1^+$ and $\hat{R}^{\gamma+}_{e}$ are the first component of
ellipticity and shear response estimated from the images with positive shear,
$\hat{e}_1^-$ and $\hat{R}^{\gamma-}_{e}$ from images with negative shear, and
$\hat{e}_1^0$ and $\hat{R}^{\gamma0}_{e}$ from undistorted images. We repeat
this whole process $250$ times with different noise realizations. For these
very well-sampled images, we expect the multiplicative bias and additive bias
are comparable on component $\hat{\gamma}_2\,$.

%% file: results.tex
In this section, we test the shear estimators derived in
Section~\ref{sec:Method} using the image simulation described in
Section~\ref{sec:sim}. The shear estimators that will be tested include the
original \cite{FPFS-Li2018} shear estimator, $\hat{\gamma}_\alpha$ (defined in
equation~\eqref{eq:estimator_0}), the shear estimator after correcting the
second-order noise bias, $\hat{\gamma}^\mathrm{A}_\alpha$
(equation~\eqref{eq:estimator_A}), the shear estimator after correcting the
selection bias from anisotropic shape noise, $\hat{\gamma}^\mathrm{B}_\alpha$
(equation~\eqref{eq:estimator_B}), and anisotropic measurement error,
$\hat{\gamma}^\mathrm{C}_\alpha$ (equation~\eqref{eq:estimator_C}). We first
test the correction for noise bias in Section~\ref{subsec:res_noisebias} and
the measurement of shape measurement error from noisy galaxy images in
Section~\ref{subsec:res_error}. Then we test the correction for selection bias
in Section~\ref{subsec:res_selbias}. Subsequently, we check the redshift
dependence of the calibration biases in Section~\ref{subsec:res_zDepBias}.
\xlrv{
Finally, we test the performance of \FPFS{} on poorly resolved galaxies in
Section~\ref{subsec:smallgal} and on stellar contamination in
Section~\ref{subsec:star}.
}

Note that we force a shear measurement measurement for every simulated galaxy
during these tests. For isolated images, the process of source detection
influences shear estimation from a population of galaxies, if that population
is determined mainly by the selection function of the detector. The right panel
of Fig.~(3) of \citet{FPFS-Li2018} showed the $s_0$ histograms of detected and
undetected galaxies in an HSC-like image simulation \citet{HSC1-GREAT3Sim} ---
most of the undetected galaxies are clustered at small $s_0\,$. Therefore, the
influence of the selection function of the detector can be removed by tuning
the lower threshold of $s_0\,$. For crowded images, removing the bias from
detection is challenging since, as shown in \citet{metaDet-Sheldon2020}, the
ability of a detection algorithm to recognize blending depends upon the
underlying shear distortion.

\begin{figure}
\begin{center}
    \includegraphics[width=0.45\textwidth]{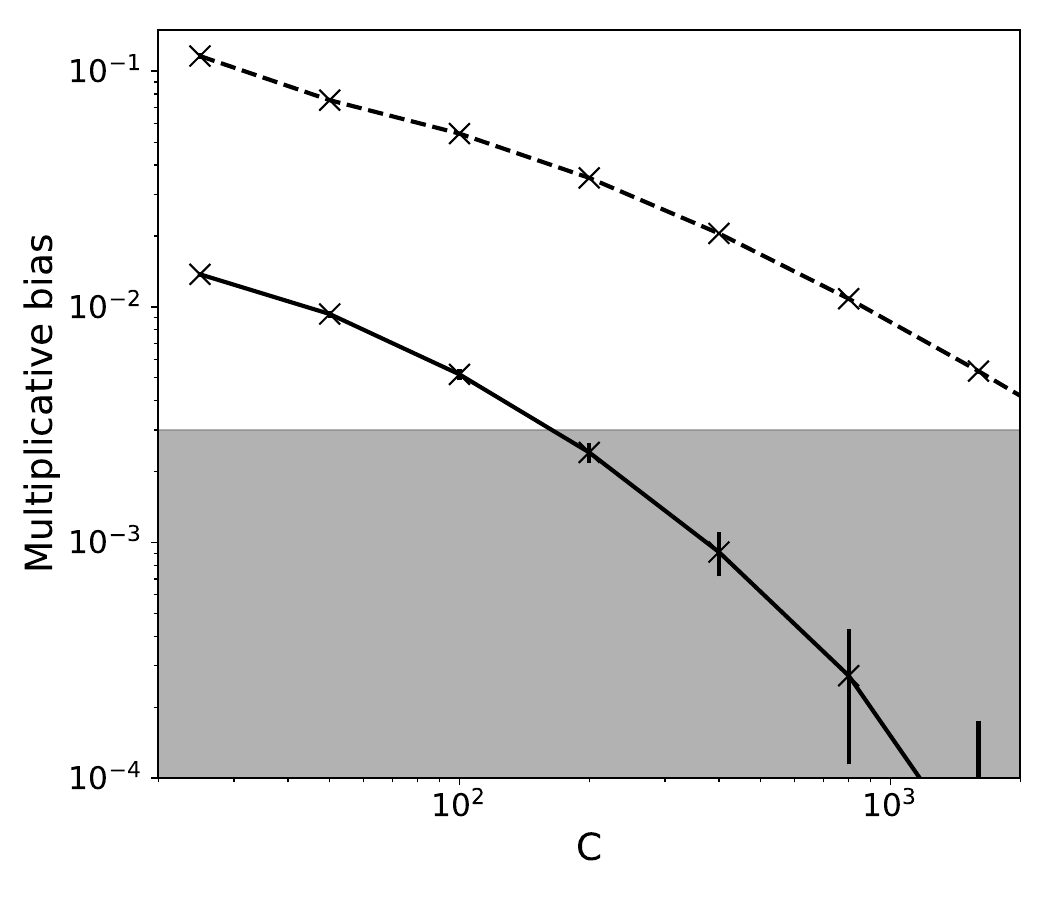}
\end{center}
\caption{
    The achieved multiplicative shear measurement bias, as a function of
    weighting parameter, $C$, both with (solid line) and without (dashed line)
    the second-order correction for nonlinear noise bias. All measured values
    of bias are negative, and their absolute values are shown. The gray region
    denotes the requirement on the control of multiplicative bias for the LSST
    surveys \citep{LSSTRequirement2018}.
    }
    \label{fig:noiBias}
\end{figure}

\subsection{Nonlinear noise bias}
\label{subsec:res_noisebias}

This subsection tests the performance of the second-order noise bias correction
derived in Section~\ref{subsec:noiseBias}. To be more specific, we change the
weighting parameter, $C$, and measure the multiplicative bias of our \FPFS{}
shear estimator with the second-order noise bias correction,
$\hat{\gamma}^\mathrm{A}$, using the simulations described in
Section~\ref{sec:sim}. The multiplicative bias in Figure~\ref{fig:noiBias} is
reduced below the requirement for the LSST survey when the weighting parameter
is greater than $200$. Since we use all the galaxies in the simulation without
any additional selection for the test shown in this subsection, selection bias
does not contribute to this result.

We also compare the result of the \FPFS{} shear estimator including the
second-order noise bias correction to that of the original \FPFS{} shear
estimator without the second-order noise bias correction in
Figure~\ref{fig:noiBias}. As shown, the noise bias is reduced by an order of
magnitude after the second-order noise bias correction. The additive bias is
constantly below $10^{-4}$, and we do not plot the additive bias here.

Note, the correction of noise bias in Section~\ref{subsec:noiseBias} assumes
that noise are homogeneous in configuration space so that noise are not
correlated in Fourier space. In our simulation described in
Section~\ref{sec:sim}, the input noise is homogeneous. In general, the
background photon noise is homogeneous in real observations; however the source
photon noise is not homogeneous, although its contributions in ground-based
surveys are small. In the presence of galaxy source photon noise, the
performance of the second-order noise bias correction is expected to be worse
than the solid line in Figure~\ref{fig:noiBias}; however, it should be better
than the dashed line Figure~\ref{fig:noiBias} that does not include any
second-order noise bias correction.

\subsection{Shape noise and shape measurement uncertainty}
\label{subsec:res_error}

This subsection calculates the statistical uncertainty in shear estimation from
a population of galaxies,
\begin{equation}
\sigma_\gamma= \frac{\sqrt{\left\langle \,\frac{1}{2}\left(\hat{e}_1^2 +\hat{e}_2^2\right)
\right\rangle}}{\hat{\res}^\gamma_e}~.
\end{equation}
This total uncertainty is a combination of noise due to galaxies' intrinsic
shapes, $e_\text{RMS}$, and shape measurement error due to realisations of
noise in images of galaxies, $\sigma_e$.  We shall assume these add in
quadrature, such that
\begin{equation}
    \label{eq:twoErrors}
    \left\langle \,\frac{1}{2}\left(\hat{e}_1^2+\hat{e}_2^2\right) \right\rangle=
    e_\text{RMS}^2 + \sigma_e^2 \,.
\end{equation}
The standard error on the mean shear measured from a population of $N$ galaxies
is thus $\sigma_\gamma/\sqrt{N}\,$. Note, however, that this value depends on
which population of galaxies it is averaged over.

\begin{figure}
\begin{center}
    \includegraphics[width=0.45\textwidth]{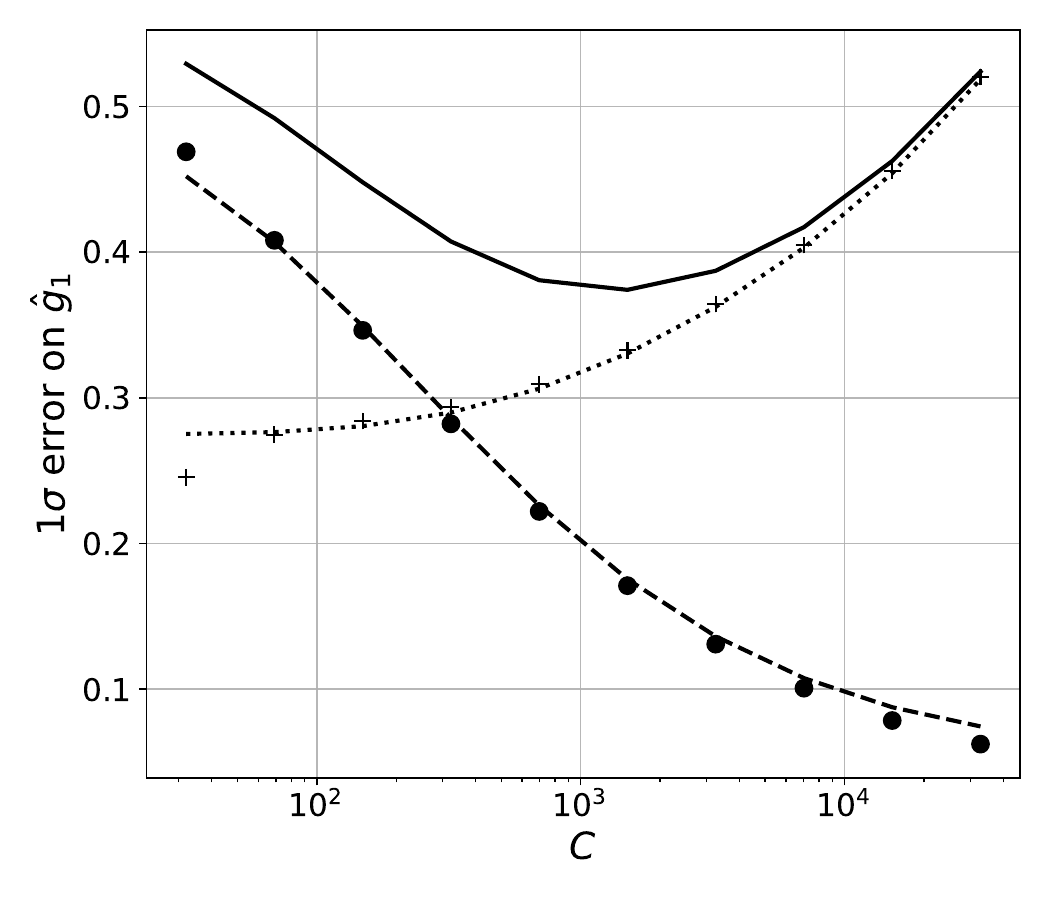}
\end{center}
\caption{
    The $1\sigma$ statistical uncertainty on shear measurements
    $\hat{\gamma}_1$ for individual galaxies (solid line), as a function of
    weighting parameter, $C\,$. The total uncertainty has contributions due to
    image noise (dashed line) and intrinsic shape noise (dotted line) --- both
    of which we measure using noiseless galaxy images that would not be
    available to a real survey. However, the `$\bullet$' (`$+$') symbols show
    the same measurement noise (intrinsic shape noise) accurately estimated
    using equation~\eqref{eq:sigmaE} and noisy galaxy images, which are
    observable.
    }
    \label{fig:errors}
\end{figure}

To obtain $e_\text{RMS}$, we measure the intrinsic \FPFS{} ellipticity of each
galaxy from a realization of the galaxy image simulation with zero noise and
zero input shear (see Section~\ref{sec:sim}). We average the two components of
ellipticity, then calculate the RMS across our sample of galaxies.  The
intrinsic shape noise, $e_\text{RMS}$, increases with weighting parameter $C$
(dotted line in Figure~\ref{fig:errors}).

To obtain $\sigma_e^2$, we first measure the total uncertainty using a noisy
realization of the galaxy image simulation (see Section~\ref{sec:sim}), again
averaging the two components of ellipticity. We then subtract $e_\text{RMS}$,
following equation~\eqref{eq:twoErrors}.  The shape measurement error,
$\sigma_e$, decreases with weighting parameter $C$ (dashed line in
Figure~\ref{fig:errors}).

\begin{figure}
\begin{center}
    \includegraphics[width=0.45\textwidth]{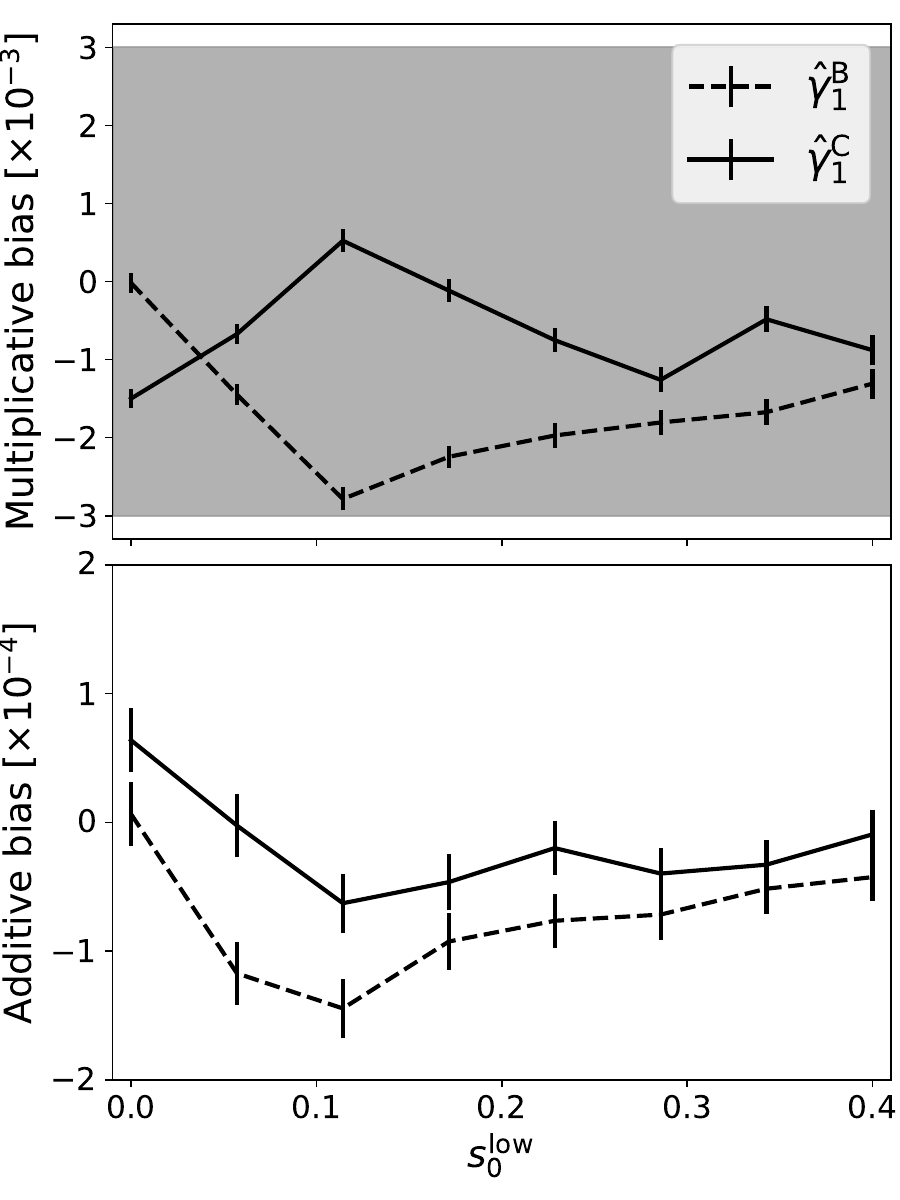}
\end{center}
\caption{
    The multiplicative bias (top panel) and additive bias (bottom panel) as a
    function of lower limit on the \FPFS{} flux ratio. The gray region
    indicates the LSST science requirement.
    }
    \label{fig:selBias}
\end{figure}

The total statistical uncertainty on $\hat{\gamma}_i$ is thus a balance between
contributions from shape noise (an increasing function of $C\,$) and from
measurement error (a decreasing function of $C$). Total uncertainty is
minimised for $1000\lesssim C\lesssim2000$, which is therefore optimal if each
galaxy in a sample is equally likely to contain shear signal. In this paper we
set $C=2000$ unless otherwise mentioned, which is close to
$2.5\,\sigma_{M_{00}}$. The corresponding nonlinear noise bias for this default
setup is well below the LSST requirement as shown in Figure~\ref{fig:noiBias}.

Shape measurement error can be also be estimated independently, using only
noisy galaxy images, and without access to noise-free versions -- as would be
required when handling real astronomical data. Following
equation~\eqref{eq:appCov2}, we estimate $\hat{\sigma}_e$ (circles in
Figure~\ref{fig:errors}), then use measurements of total noise and
equation~\eqref{eq:twoErrors} to estimate intrinsic shape noise
$\hat{e}_\text{RMS}$ (crosses in Figure~\ref{fig:errors}). These reproduce the
measurements from noiseless image simulations with remarkable accuracy.

The total \FPFS{} shear measurement uncertainty is similar to that from the
calibrated \reGauss{} shear estimator \citet{HSC1-GREAT3Sim}. To demonstrate
this, we run the HSC pipeline (\hscPipe{7}) for source detection and shape
measurement on our simulated images (see \citealt{HSC1-pipeline} for details on
the pipeline), which includes catalogue cuts at\,\ $i<24.5$,
$\mathrm{resolution}>0.3$, and $\mathrm{SNR}>10$ \citep{HSC1-catalog}.  To
weight the galaxies, we use fixed $C=2000$ for \FPFS{}. For \reGauss{}, we use
the optimal weight of a real galaxy in the first-year HSC shear catalogue,
selected as the closest match in the $\log\left( \mathrm{SNR}
\right)$-$\log\left( \mathrm{resolution} \right)$ plane. Since the \reGauss{}
algorithm is subject to certain forms of shear estimation bias (e.g.\ model
bias, noise bias), we also use the \reGauss{} ellipticities measured from the
simulation with $\gamma_1=0.02$ and $\gamma_1=-0.02$ to linearly calibrate its
shear response. For this galaxy sample, which has higher S/N than the previous
sample, we find shear estimation uncertainty of $\sigma_\gamma=0.298$ for
\FPFS{}, and $\sigma_\gamma=0.288$ for \reGauss{}.

\subsection{Selection bias}
\label{subsec:res_selbias}

\begin{figure}
\begin{center}
    \includegraphics[width=0.45\textwidth]{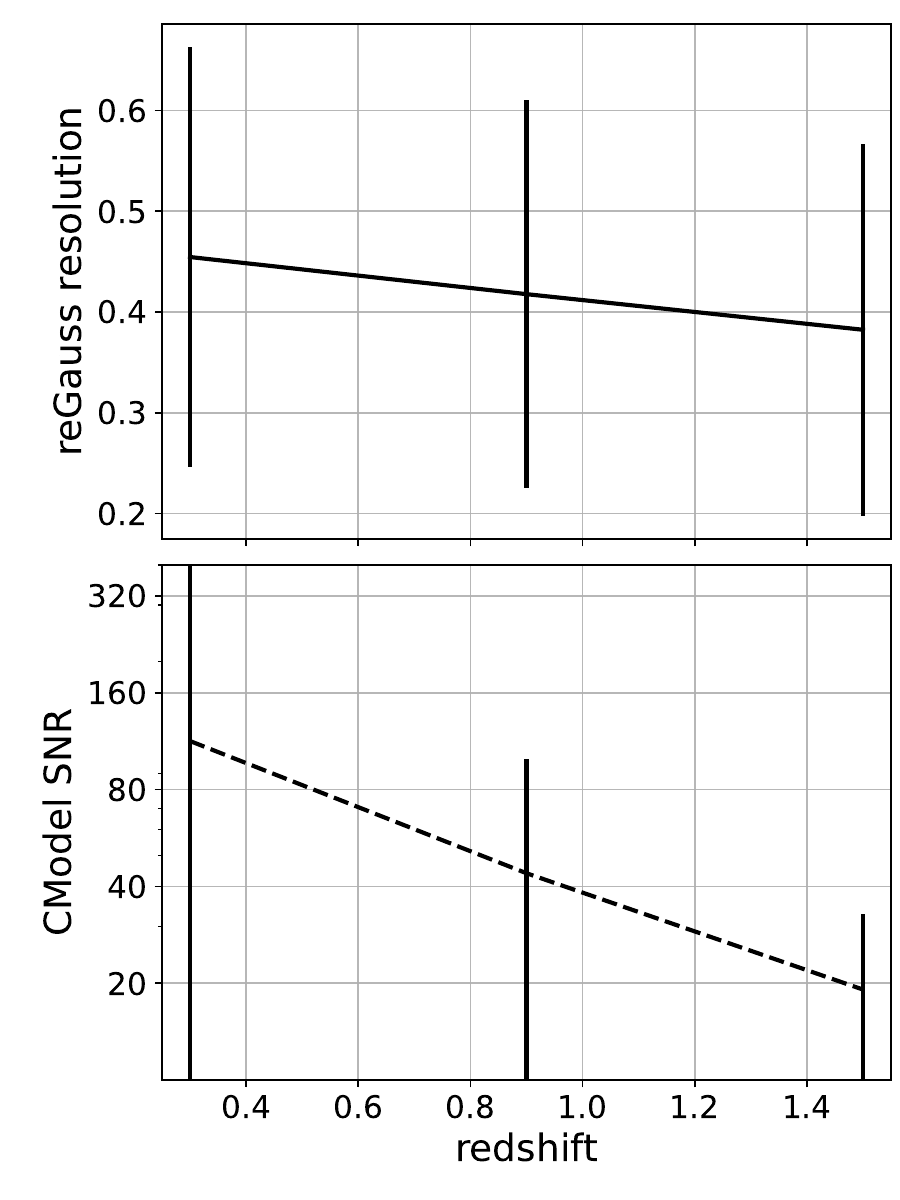}
\end{center}
\caption{
    Average \reGauss{} resolution (top panel) and CModel SNR (bottom panel) as
    functions of redshift. The errorbas show the $1\sigma$ scatter of the
    corresponding observables in each bin.
    }
    \label{fig:zDependsnrRes}
\end{figure}

This subsection tests the performance of the selection bias correction.
Specifically, we adjust the faint-end cut on the \FPFS{} flux ratio,
$\hat{s}_0>s_0^\mathrm{low}$, and estimate the shear measurement bias in shear
estimators $\hat{\gamma}^\mathrm{B}_\alpha$ (equation~\eqref{eq:estimator_B}) and
$\hat{\gamma}^\mathrm{C}_\alpha$ (equation~\eqref{eq:estimator_C}). Throughout
this section, $s^{\mathrm{upp}}=\infty$, and the weighting parameter is set to
$C=2000\,$.

The measured multiplicative biases (top panel of Figure~\ref{fig:selBias}) are
within the LSST science requirement, for estimators both with
($\hat{\gamma}^\mathrm{C}$) and without ($\hat{\gamma}^\mathrm{B}$) correction
for selection bias due to anisotropic measurement error.  Estimator
$\hat{\gamma}^\mathrm{C}$ improves upon $\hat{\gamma}^\mathrm{B}$ by
$\sim$$10^{-3}$ on average, which indicates that the multiplicative bias due to
anisotropic measurement error is at about this level.

The measured additive biases (bottom panel of Figure~\ref{fig:selBias}) are
below $1.5\times 10^{-4}$ for $\hat{\gamma}^\mathrm{B}$ and below $1\times
10^{-4}$ for $\hat{\gamma}^\mathrm{C}$, which indicates that the additive bias
due to the anisotropic measurement error is at the level of $10^{-4}\,$.

\subsection{Redshift dependence of bias}
\label{subsec:res_zDepBias}

\begin{figure}
\begin{center}
    \includegraphics[width=0.45\textwidth]{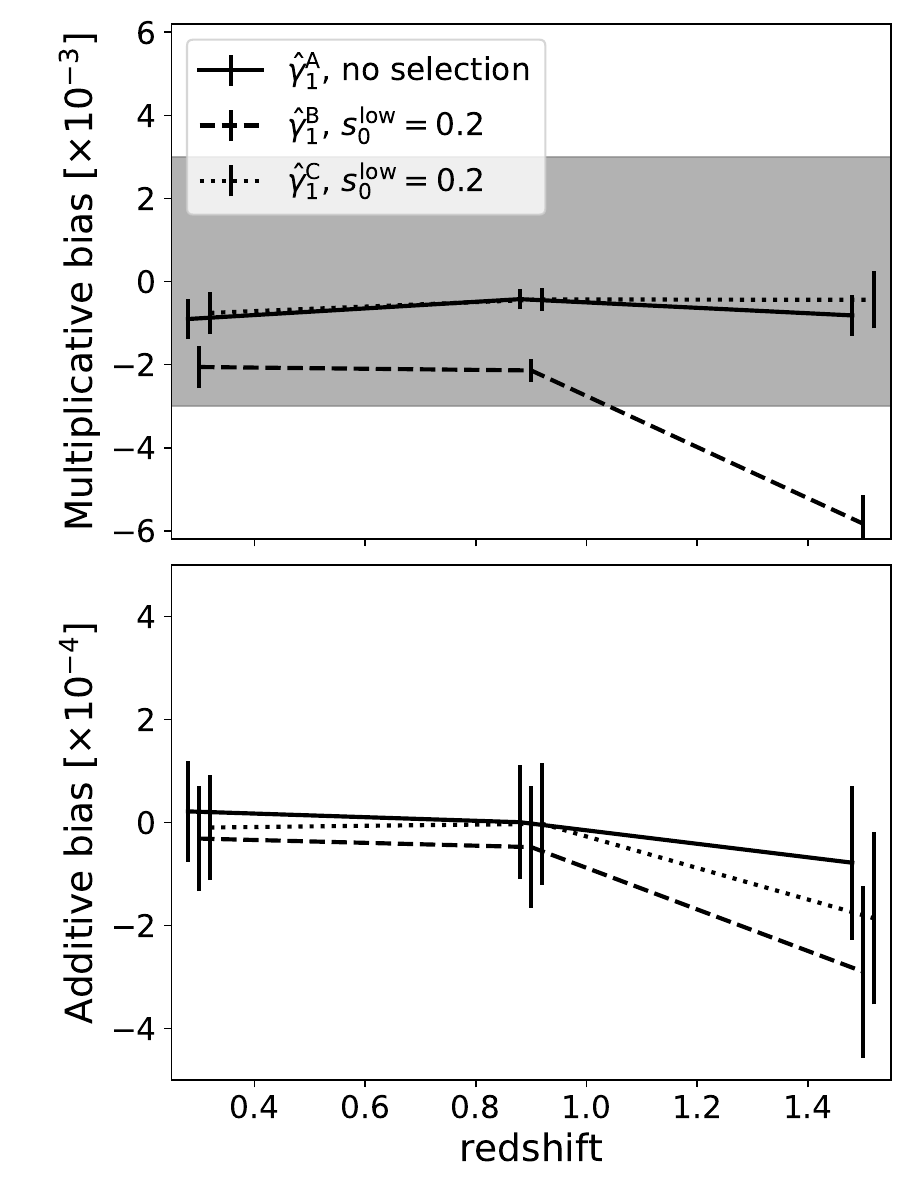}
\end{center}
\caption{
    Shear measurement multiplicative bias (top panel) and additive bias (bottom
    panel), as a function of galaxy redshift. The gray region indicates the
    LSST science requirement. Plotted points are offset by $\pm 0.02$ to
    prevent the error bars from overlapping.
    }
    \label{fig:zDepend}
\end{figure}

\xlrv{
This subsection tests whether the shear measurement biases depend upon galaxy
redshift. We divide simulated galaxies into three bins ($0\leq z<0.6$,
$0.6\leq z<1.2$, $1.2\leq z<1.8$) according to the photometric redshift of the
input COSMOS galaxies \citep{COSMOSZ_Ilbert2009}. The average \reGauss{}
resolution and CModel SNR as functions of redshift are shown in
Figure~\ref{fig:zDependsnrRes}.
}

First, we test shear estimator $\hat{\gamma}^\mathrm{A}$ without selecting by
any observables other than the COSMOS redshift. Since the COSMOS redshifts are
from input galaxies and are not influenced by shear distortion or image noise,
the redshift binning does not lead to selection bias. In addition,
$\hat{\gamma}^\mathrm{A}$ does not account for selection bias; therefore, this
test measures any redshift-dependence of the nonlinear noise bias.  We find
multiplicative bias $|m|<1\times10^{-3}\,$, and additive bias
$|c|<1\times10^{-4}\,$ at all redshifts (solid lines in
Figure~\ref{fig:zDepend}).

Second, we test shear estimator $\hat{\gamma}^\mathrm{B}$ on a galaxy sample
selected with $\hat{s}_0>s_0^\mathrm{low}=0.2$. The estimator does not account
for the selection bias due to anisotropic measurement error, so this test
isolates the performance of its correction for selection bias due to
anisotropic shape noise. We find multiplicative bias $|m|\approx2\times
10^{-3}$ for redshift $z<1$, increasing to $6\times 10^{-3}$ at high redshift
$1.2 \leq z< 1.8$. The additive bias is $2\sigma$ consistent with zero at all
redshifts (dashed lines in Figure~\ref{fig:zDepend}).

Finally, we test shear estimator $\hat{\gamma}^\mathrm{C}$ on the same galaxy
sample with $\hat{s}_0>s_0^\mathrm{low}=0.2$. This estimator accounts for
selection bias due to both anisotropic shape noise and anisotropic measurement
error. It produces multiplicative bias consistently $<1\times10^{-3}$, and
additive bias that is $2\sigma$ consistent with zero (dotted lines in
Figure~\ref{fig:zDepend}). Comparing the multiplicative biases of
$\hat{\gamma}^\mathrm{B}$ and $\hat{\gamma}^\mathrm{C}$, we find that the
amplitude of the selection bias due to the anisotropic measurement error is a
few part in $10^{3}\,$. For these isolated galaxies, our final \FPFS{} shear
estimator meets the science requirement of the LSST survey, shown as a gray
region in Figure~\ref{fig:zDepend}.

\subsection{Performance with very small source galaxies}
\label{subsec:smallgal}

\xlrv{
Since our fiducial image simulation is based on a training sample of galaxies
resolved by HST and with magnitude $i<25.2$, it does not include the smallest
galaxies that a future survey might have ambition to measure. To test the
accuracy of \FPFS{} on galaxies that are barely resolved (especially by
ground-based observations), we use \galsim{} to simulate small galaxies
composed of $20$ points randomly distributed \citep{Z15} to follow a $2$D
Gaussian profile with input half light radius ranging from $0\farcs07$ to
$0\farcs2\,$.
The flux of each knot is the same.
The measured \reGauss{} resolutions range from $0.12$ to $0.27$ as shown in
Table~\ref{tab:small_gals}. Each galaxy sample has $4\times10^7$ galaxies and
with an average CModel SNR $\sim15\,$, and each galaxy is rotated by $45\deg$
four times to reduce shape noise from both spin $m=2$ and spin $m=4$
quantities. Here we do not add any additional selection, so that selection bias
is not present. As shown in Table~\ref{tab:small_gals}, \FPFS{} can accurately
measure shear from even extremely small and faint galaxies.
If mixed with bigger and brighter galaxies, they will receive a low weight and
contribute little to the signal. Crucially, they will not bias it.
}

\begin{table}
\caption{
The average \reGauss{} resolution (first row), shear multiplicative bias
(second row), and shear additive bias (third row) for three samples of galaxies
simulated as a collection of random point sources. All these galaxies are
smaller than those in the HSC shape catalog, which analyses only galaxies with
\reGauss{} resolution greater than $0.3$ \citep{HSC3-catalog}.
}
\begin{center}
\begin{tabular}{c| c c c}
                        & smallest      & smaller       & small \\
\hline
\reGauss{} resolution   & 0.12          & 0.19          & 0.27          \\
$m_1(10^{-4})$          & $0.4 \pm 0.9$ & $-1.4 \pm 0.7$& $-1.0\pm 0.5$ \\
$c_1(10^{-4})$          & $-2.8\pm 4.5$ & $-0.8\pm 1.1$ & $-0.5\pm 0.7$ \\
\end{tabular}
\end{center}
\label{tab:small_gals}
\end{table}

\subsection{Stellar contamination}
\label{subsec:star}

\begin{figure}
\begin{center}
    \includegraphics[width=0.45\textwidth]{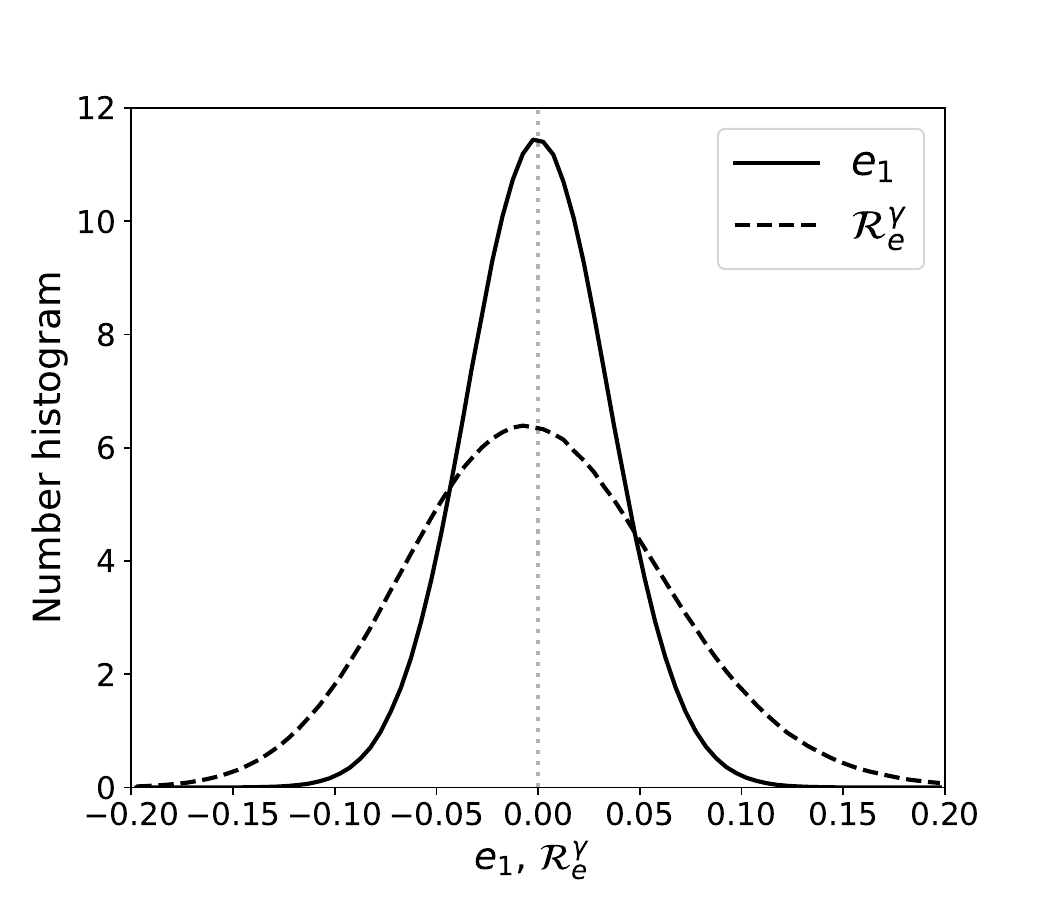}
\end{center}
\caption{
    Normalized number histograms of the first component of \FPFS{} ellipticity
    $e_1$ (solid line) and \FPFS{} response $\res_e^\gamma$ (dashed line)
    measured from simulated images of stars. The vertical dotted line shows the
    expectation value.
    }
    \label{fig:stellarERHist}
\end{figure}

\xlrv{
In real observations, stars may not be perfectly removed from the galaxy
sample. To test the performance of our \FPFS{} shear estimator with such
stellar contamination, we simulate $1\times10^7$ star images with an average
SNR $17.8\,$. This corresponds to an extreme situation that the object's
\reGauss{} resolution equals zero. Then we measure the \FPFS{} ellipticity and
\FPFS{} response from these stars. Again, we neglect the PSF model errors and
assume that we know the two-point correlation function of noise.

Measurements stars yield mean values $\langle\hat{e}_1\rangle = (1.0\pm 1.1)
\times10^{-5}$ and $\res_e^\gamma=(1.8 \pm 2.0) \times10^{-5}$
(Figure~\ref{fig:stellarERHist}). That the expectation value of both is
consistent with zero (and much smaller than the mean response of simulated HST
COSMOS galaxies, $\abs{\res_e^\gamma}\sim0.18$), ensures that \FPFS{} is robust
to stellar contamination, so long as the PSF is well determined. Stellar
contamination of $n$\% will reduce the numerator and denominator of
equation~\eqref{eq:estimator_C} by $n$\%, leaving the shear estimator unbiased.
This is the same good property as \metacal{}, demonstrated in Figure~5 of
\citet{metacal-Sheldon2017}.
}

%% file: appendix.tex
\section{Second-order revision for nonlinear noise bias}
\label{app:noiseBias}

Here we present the expectation values of noisy, measurable quantities
(indicated with a tilde), relative to those of the unobservable, noiseless
quanties (without a tilde). We only keep to the second-order terms of noise
residuals and neglect the higher-order terms. First, we obtain the expectation
for $\tilde{s}_0$ and $\tilde{s}_4$:
\begin{equation}\label{eq:appCov1}
    \begin{split}
    \left\langle \tilde{s}_0 \right\rangle
    &=\left\langle
        s_0
        \left(1+
        \frac{\mathcal{V}_{0000}}
        {(M_{00}+C)^2}
    \right)\right\rangle\\
    &-\left\langle
    \frac{\mathcal{V}_{0000}}
    {(M_{00}+C)^2}
    \right\rangle,\\
    \left\langle \tilde{s}_4 \right\rangle
    &=\left\langle
        s_4
        \left(1+
        \frac{\mathcal{V}_{0000}}
        {(M_{00}+C)^2}
    \right)\right\rangle\\
    &-\left\langle
    \frac{\mathcal{V}_{0040}}
    {(M_{00}+C)^2}
    \right\rangle\,.
    \end{split}
\end{equation}

Then we use the covariance matrix of the shapelet modes
(equation~\eqref{eq:shapeletcov}) to derive the expectation for
$\tilde{e}_{1,2}^2 $, $\tilde{s}_0^2$, and $\tilde{e}_{1,2}\tilde{s}_0$:
\begin{equation}\label{eq:appCov2}
    \begin{split}
    \left\langle \tilde{e}_1^2\right\rangle
    &= \left\langle e_1^2\left(1+3
        \frac{\mathcal{V}_{0000}}
            {(M_{00}+C)^2}
        \right) \right\rangle\\
    &+ \left\langle
        \frac{\mathcal{V}_{22c22c}}{(M_{00}+C)^2}
    - 4e_1\frac{\mathcal{V}_{0022c}}
        {(M_{00}+C)^2}
        \right\rangle,\\
    \left\langle \tilde{e}_2^2\right\rangle
    &= \left\langle e_2^2\left(1+3
        \frac{\mathcal{V}_{0000}}
            {(M_{00}+C)^2}
        \right) \right\rangle\\
    &+ \left\langle
        \frac{\mathcal{V}_{22s22s}}{(M_{00}+C)^2}
    - 4e_2\frac{\mathcal{V}_{0022s}}
        {(M_{00}+C)^2}
        \right\rangle,\\
    \left\langle \tilde{s}_0^2\right\rangle
    &= \left\langle s_0^2\left(1+3
        \frac{\mathcal{V}_{0000}}
            {(M_{00}+C)^2}
        \right) \right\rangle\\
    &+ \left\langle
        \frac{\mathcal{V}_{0000}}{(M_{00}+C)^2}
    - 4s_0\frac{\mathcal{V}_{0000}}
        {(M_{00}+C)^2}
    \right\rangle,\\
    \left\langle \tilde{e}_1 \tilde{s}_0 \right\rangle
    &= \left\langle e_1 s_0 \left(1+3
        \frac{\mathcal{V}_{0000}}
            {(M_{00}+C)^2}
        \right) \right\rangle\\
    &+ \left\langle
        \frac{\mathcal{V}_{0022c}}{(M_{00}+C)^2}
    - 2s_0\frac{\mathcal{V}_{0022c}}
        {(M_{00}+C)^2}
    \right\rangle\\
    &-\left\langle
     2e_1\frac{\mathcal{V}_{0000}}
        {(M_{00}+C)^2}
    \right\rangle,\\
    \left\langle \tilde{e}_2 \tilde{s}_0 \right\rangle
    &= \left\langle e_2 s_0 \left(1+3
        \frac{\mathcal{V}_{0000}}
            {(M_{00}+C)^2}
        \right) \right\rangle\\
    &+ \left\langle
        \frac{\mathcal{V}_{0022s}}{(M_{00}+C)^2}
    - 2s_0\frac{\mathcal{V}_{0022s}}
        {(M_{00}+C)^2}
    \right\rangle\\
    &-\left\langle
     2e_2\frac{\mathcal{V}_{0000}}
        {(M_{00}+C)^2}
    \right\rangle\,.
    \end{split}
\end{equation}
\xlrv{
These quantities are used to derive the variance of measurement error on
\FPFS{} ellipticity $\delta{e}_{1,2}$ defined in equation~\eqref{eq:deltaElli}
and flux ratio $\delta{s}_0$ defined in equation~\eqref{eq:deltaS0} due to the
photon noise on galaxy images where noise terms of fourth order and higher are
neglected.
\begin{equation}\label{eq:app_dede}
    \begin{split}
    \langle (\delta{e}_1)^2 \rangle
    &= \left \langle \hat{e}_1^2 - e_1^2 \right\rangle\\
    &= \left\langle \tilde{e}_1^2\, \frac{\mathcal{V}_{0000}}{(\tilde{M}_{00}+C)^2}
    + \frac{\mathcal{V}_{22c22c}}{(\tilde{M}_{00}+C)^2}
    -2\,\tilde{e}_1\,\frac{\mathcal{V}_{0022c}}{(\tilde{M}_{00}+C)^2} \right\rangle\,,\\
    \langle (\delta{e}_2)^2 \rangle
    &= \left\langle \hat{e}_2^2 - e_2^2 \right\rangle\\
    &= \left\langle \tilde{e}_2^2\, \frac{\mathcal{V}_{0000}}{(\tilde{M}_{00}+C)^2}
    + \frac{\mathcal{V}_{22s22s}}{(\tilde{M}_{00}+C)^2}
    - 2\,\tilde{e}_2\,\frac{\mathcal{V}_{0022s}}{(\tilde{M}_{00}+C)^2} \right\rangle\,,\\
    \langle (\delta{s}_0)^2 \rangle
    &= \left \langle \hat{s}_0^2 - s_0^2 \right\rangle\\
    &= \left\langle \tilde{s}_0^2\frac{\mathcal{V}_{0000}}{(\tilde{M}_{00}+C)^2}
    + \frac{\mathcal{V}_{0000}}{(\tilde{M}_{00}+C)^2}
    - 2\,\tilde{s}_0\,\frac{\mathcal{V}_{0000}}{(\tilde{M}_{00}+C)^2} \right\rangle\,.\\
    \end{split}
\end{equation}
In addition, the correlation between the measurement errors $\delta{e}_{1,2}$
and $\delta{s}_0$ is given by
\begin{equation}\label{eq:app_deds}
    \begin{split}
    \langle \delta{s}_0 \delta{e}_1 \rangle
    &= \left\langle \hat{s}_0\hat{e}_1 - s_0e_1 \right\rangle\\
    &= \left\langle \tilde{e}_1\tilde{s}_0 \frac{\mathcal{V}_{0000}}{(\tilde{M}_{00}+C)^2}
    + \frac{\mathcal{V}_{0022c}}{(\tilde{M}_{00}+C)^2} \right\rangle\\
    &- \left\langle \tilde{s}_0\,\frac{\mathcal{V}_{0022c}}{(\tilde{M}_{00}+C)^2}
    + \tilde{e}_1\,\frac{\mathcal{V}_{0000}}{(\tilde{M}_{00}+C)^2}\right\rangle,\\
    \langle \delta{s}_0 \delta{e}_2 \rangle
    &= \left\langle \hat{s}_0\hat{e}_2 - s_0e_2 \right\rangle\\
    &= \left\langle \tilde{e}_2\tilde{s}_0 \frac{\mathcal{V}_{0000}}{(\tilde{M}_{00}+C)^2}
    + \frac{\mathcal{V}_{0022s}}{(\tilde{M}_{00}+C)^2} \right\rangle\\
    &- \left\langle \tilde{s}_0\,\frac{\mathcal{V}_{0022s}}{(\tilde{M}_{00}+C)^2}
    + \tilde{e}_2\,\frac{\mathcal{V}_{0000}}{(\tilde{M}_{00}+C)^2}\right\rangle,\\
    \end{split}
\end{equation}
}

Finally we derive expectation for the noisy quantities related to the selection
shear response:
\begin{equation}\label{eq:appCov3}
    \begin{split}
    \left\langle (\tilde{e}_1)^2 \tilde{s}_0\right\rangle
    &= \left\langle (e_1)^2 s_0
    \left( 1 + 6\frac{\mathcal{V}_{0000}} {(M_{00}+C)^2} \right)
    \right\rangle\\
    &-3\left\langle (e_1)^2\frac{\mathcal{V}_{0000}} {(M_{00}+C)^2}
    \right\rangle
    + \left\langle s_0\frac{\mathcal{V}_{22c22c}} {(M_{00}+C)^2}
    \right\rangle\\
    &+\left\langle 2e_1\left(1-3s_0\right)
        \frac{\mathcal{V}_{0022c}} {(M_{00}+C)^2}
    \right\rangle,\\
    \left\langle (\tilde{e}_2)^2 \tilde{s}_0\right\rangle
    &= \left\langle (e_2)^2 s_0
    \left( 1 + 6\frac{\mathcal{V}_{0000}} {(M_{00}+C)^2} \right)
    \right\rangle\\
    &-3\left\langle (e_2)^2\frac{\mathcal{V}_{0000}} {(M_{00}+C)^2}
    \right\rangle
    + \left\langle s_0\frac{\mathcal{V}_{22s22s}} {(M_{00}+C)^2}
    \right\rangle\\
    &+\left\langle 2e_2\left(1-3s_0\right)
        \frac{\mathcal{V}_{0022s}} {(M_{00}+C)^2}
    \right\rangle\,.\\
    \end{split}
\end{equation}